\documentclass[12pt, a4paper, twoside, notitlepage]{report}

\usepackage[dvipdf]{graphicx}
\usepackage{amsmath}
\usepackage{colordvi}
\usepackage{hyperref}
\usepackage{SIunits}
\usepackage{here}
\usepackage{fancyhdr}
\usepackage[polish,english]{babel}

\newcommand\keV{\kilo\electronvolt}
\newcommand\MeV{\mega\electronvolt}
\newcommand\geant{\textsc{Geant4}\/}
\newcommand\link[2]{\href{#1}{\scriptsize{\Blue{\underline{\fontfamily{phv}\selectfont #2}}}}}
\newcommand\email[2]{\href{#1}{\Blue{\fontfamily{pzc}\selectfont #2}}}

\newcommand\iu{\operatorname{i}}        

\def\twomatrix#1#2#3#4{{\left(\begin{array}{cc}{#1}&{#2}\\{#3}&{#4}\end{array}\right)}}

\def\pspint{{\twomatrix1{\,\,0}0{\,\,\,1}}}
\def\pspinx{{\twomatrix0{\,\,\,1}1{\,\,\,0}}}
\def\pspiny{{\twomatrix0{\!-\!\iu}\iu{\,\,\,0}}}
\def\pspinz{{\twomatrix1{\,\,0}0{\!-\!1}}}

\long\def\dedication#1{\gdef\@dedication{#1}}
\long\def\acknowledgements#1{\gdef\@acknowledgements{#1}}
\def\@dedication{\@latex@warning@no@line{No [optional]
    \noexpand\dedication provided}}
\def\@acknowledgements{\@latex@warning@no@line{No [optional]
    \noexpand\acknowledgements given}}
\newcommand{\makededication}%
  {\chapter{Dedication}
   \@dedication \par%
   \global\let\makededication\relax
   \global\let\dedication\relax
   \global\let\@dedication\@empty
  }

\newcommand{\makeacknowledgements}%
  {\chapter{Acknowledgements}
   \@acknowledgements \par%
   \global\let\makeacknowledgements\relax
   \global\let\acknowledgements\relax
   \global\let\@acknowledgements\@empty
}

\begin{document}
\pagestyle{empty}
\begin{titlepage}
  \title{Analysis of experimental uncertainties in the $R$-correlation measurement in the decay of free neutrons}
  \author{Marcin Ku\'zniak}
\end{titlepage}
\thispagestyle{empty}

\begin{center}

{\Huge\    JAGELLONIAN UNIVERSITY         \\ 
            INSTITUTE OF PHYSICS          \\
}
\vspace{2.75cm}
\begin{figure}[h]
  \centering
\end{figure}
\vspace{2.75cm}
{\bf{\LARGE Analysis of experimental uncertainties in the $R$-correlation measurement in the decay of free neutrons
}}

\vspace{2.5cm}

{\large                 Marcin Ku\'zniak} \\
\email{mailto:kuzniak@if.uj.edu.pl}{kuzniak@if.uj.edu.pl}
\vspace{5.0cm}

\end{center}

\vspace{1.5cm}

\begin{flushright}
Master's Thesis prepared at the Nuclear Physics Department\\
guided by prof. dr hab. Kazimierz Bodek  \\
\end{flushright}
\vspace{0.5cm}
\begin{center}
    Krak\'ow 2004 
\end{center}

\cleardoublepage
\begin{abstract}
The experiment aiming at the simultaneous determination of the two transversal polarisation components of electrons emitted in the decay 
of free, polarised neutrons is in progress at the Paul Scherrer Institute (Villigen, Switzerland). The non-zero value of $R$ coefficient,
proportional to the 
polarisation component, which is perpendicular to the plane spanned by the spin of the decaying neutron and the electron momentum, would 
prove a violation of time reversal symmetry and thus physics beyond the Standard Model. The planned accuracy of the measurement is 
of order 0.005. To reach this value, the systematic effects in the experiment have to be controlled on a similar level of accuracy. 
The emphasis of this master's thesis 
is put on the search of systematic effects by the means of dedicated Monte Carlo simulation, based on extended 
\geant\ package. Implementation details are discussed and the new added features are tested. Finally, the $\beta$ decay asymmetry 
induced systematic effect, resulting in false contribution to $R$-coefficient is recognised and investigated.
\end{abstract}
\selectlanguage{polish}
\begin{abstract}
W Instytucie Paula Scherrera (Villigen, Szwajcaria) prowadzony jest eksperyment maj\aob cy na celu jednoczesny pomiar obu poprzecznych 
sk\l adowych polaryzacji 
elektron\'ow emitowanych w rozpadzie swobodnych, spolaryzowanych neutron\'ow. Niezerowa warto\'s\'c wsp\'olczynnika $R$, proporcjonalnego
do tej sk\l a\-do\-wej polaryzacji, kt\'ora jest prostopad\l a do p\l aszczyzny tworzonej przez spin neutronu i p\eob d elektronu, 
\'swiadczy\l aby o istnieniu procesu \l ami\aob cego symetri\eob\ wzgl\eob dem odwr\'ocenia czasu, wykraczaj\aob cego poza obecnie 
uznawan\aob\  struktur\eob\  Modelu Standardowego. Aby m\'oc osiagn\aob\'c planowan\aob\ dok\l adno\'s\'c pomiaru (0.005), 
konieczne jest oszacowanie mo\.zliwych efekt\'ow systematycznych. W poni\.zszej pracy magisterskiej nacisk po\l o\.zono na 
poszukiwanie efekt\'ow systematycznych przy pomocy specjalnie do tego celu stworzonej symulacji Monte Carlo, opartej na rozszerzonym 
pakiecie \geant. Om\'owione s\aob\ szczeg\'oly
implementacji oraz wyniki test\'ow nowych, dodanych do pakietu funkcji. Opisano te\.z i przeanalizowano odkryty efekt systematyczny, 
spowodowany asymetri\aob\ rozpadu $\beta$ i wnosz\aob cy fa\l szywy wk\l ad do mierzonego wsp\'olczynnika $R$.
\end{abstract}
\selectlanguage{english}
\cleardoublepage
\chapter*{Acknowledgements}
The work presented in this thesis would not have been possible without the involvement of a number of people. 
I would like to thank the following persons in particular:
\begin{itemize}
\item Prof. Kazimierz Bodek, my supervisor, for his preciuos advice and suggestions concerning my work, for great knowledge and 
  experience he shares with his students. I would also like to thank for involving me in physics of cold neutrons and for giving 
  me the possibility to work with the nTRV group;
\item dr Stanis\l aw Kistryn for being my tutor during two first years of my studies, for his help and sense of humour;
\item Prof. Reinhard Kulessa for allowing me to prepare this thesis in the Nuclear Physics Department of the Jagellonian University;
\item Prof. Bogus\l aw Kamys for interesting lectures and a lot of help;
\item dr El\.zbieta Stephan for help and ideas concerning the Monte Carlo simulation;
\item dr Adam Kozela, dr Jacek Zejma, Aleksandra Bia\l ek, Pierre Gorel and Jacek Pulut, other members of our group, for constant 
  assistance and lots of fruitful discussions;
\item all residents, guests and numerous friends of the famous students' flat on Zam\-ko\-wa Street for lots of discussions;
\item my colleagues Ma\l gorzata Kasprzak, Joanna Przerwa, Micha\l\ Janusz and Tytus Smoli\'nski for the great atmosphere of daily
  work;
\item my dear Parents for their love, enormous patience and permanent support during the five years of my studies.
\end{itemize}
\cleardoublepage
\tableofcontents
\cleardoublepage
\pagestyle{headings}
\chapter{Introduction}
From the nuclear physicist's point of view next few years are going to be really interesting.
On the one hand, the Standard Model (SM) has still an impressive predictive power and so far
in particle physics there has been a total agreement with experimental evidence. On the other hand,
SM is not believed to be a complete theory. Nowadays there are plenty of much more elegant
new theories or at least extensions of the SM. Moreover, there is one huge discrepancy between
its predictions and astronomical observations, namely the baryon asymmetry of the Universe 
which is
by a factor $10^{8}$ larger than expected. In other words,
according to the SM, large concentrations of baryon matter, such as galaxies, could not have even
existed and the Universe should have been more like a uniform ``sea'' of radiation.

One of very few reasonable explanations of this paradox requires the breaking of combined charge
conjugation and parity symmetry ($\cal CP$). Assuming that $\cal CPT$ symmetry (combined
time-reversal symmetry $\cal T$ and $\cal CP$) is conserved, what is required by all
renormalisable quantum field theories, $\cal CP$ violation implies $\cal T$ violation.
Both $\cal CP$ or $\cal T$-violating processes have been so far observed only in the neutral
$K$ and $B$ meson systems and their mechanism has been already included in the SM by
introducing a quark mixing mechanism. However, to explain the problem of baryon asymmetry
additional sources of $\cal CP$ or $\cal T$ breaking have to be found.

Measurements of vector correlations in particle decays and searching for electric dipole moments
of particles belong to the most promising ways to discover new $\cal T$-violating processes
not predicted by the SM. For both types of experiments, physics of cold neutrons is
especially interesting, due to the availability of high intensity polarised beams. The
experiment aimed at the determination of the $R$-correlation parameter (mixed product of
neutron spin, electron momentum and electron spin) in the $\beta^-$ decay of
free, polarised cold neutrons will start data taking this summer at the Paul Scherrer Institute
(Villigen, Switzerland). It is going to be the first such a measurement for the decay of free
neutrons, bearing a potential to detect either a non-standard value, inconsistent with zero or
to provide important constraints for the $\cal T$-violating scalar and tensor couplings in the
semileptonic weak interactions.
\\\\
The subject of this thesis is the analysis of experimental uncertainties in the $R$-correlation
experiment. The measurement is planned to achieve accuracy of 0.005, with the main contribution
to the error coming from the counting statistics. In order to perform detailed study of possible
{\em systematic} effects, a dedicated Monte Carlo simulation (based on \geant\ package) has
been created and exploited.
\\\\
Before the main goal could have been completed, the following
intermediate steps had to be fulfilled:
\begin{itemize}
\item Unification of all existing parts of the source code in one general simulation of the
  whole experiment
\item Implementation of new geometry of the experimental setup
\item Modification of \geant\ libraries to include electron polarisation transport and the
  spin dependent neutron $\beta^-$ decay
\item Further software development and optimisation
\item Actual simulations and data analysis
\end{itemize}

The next chapter briefly describes the theoretical aspects of the
neutron $\beta^-$ decay, which are essential for understanding the
principle of the experiment. In addition, interactions of
electrons with matter and  foundations of the polarisation theory
are sketched, especially the formulae used in the simulation. The
third chapter contains detailed information about the measurement
itself and the description of the experimental apparatus. In
further chapters, the Monte Carlo simulation and its results are
presented and discussed. Appendices contain practical
information on the software usage.
\cleardoublepage

\chapter{Theory}
\section{The neutron \texorpdfstring{$\beta^-$ }{beta- }decay and angular correlations}\label{sec-ntheory}
\begin{figure}[!h]
  \label{fig-beta}
  \center
  \includegraphics[width=0.49\textwidth]{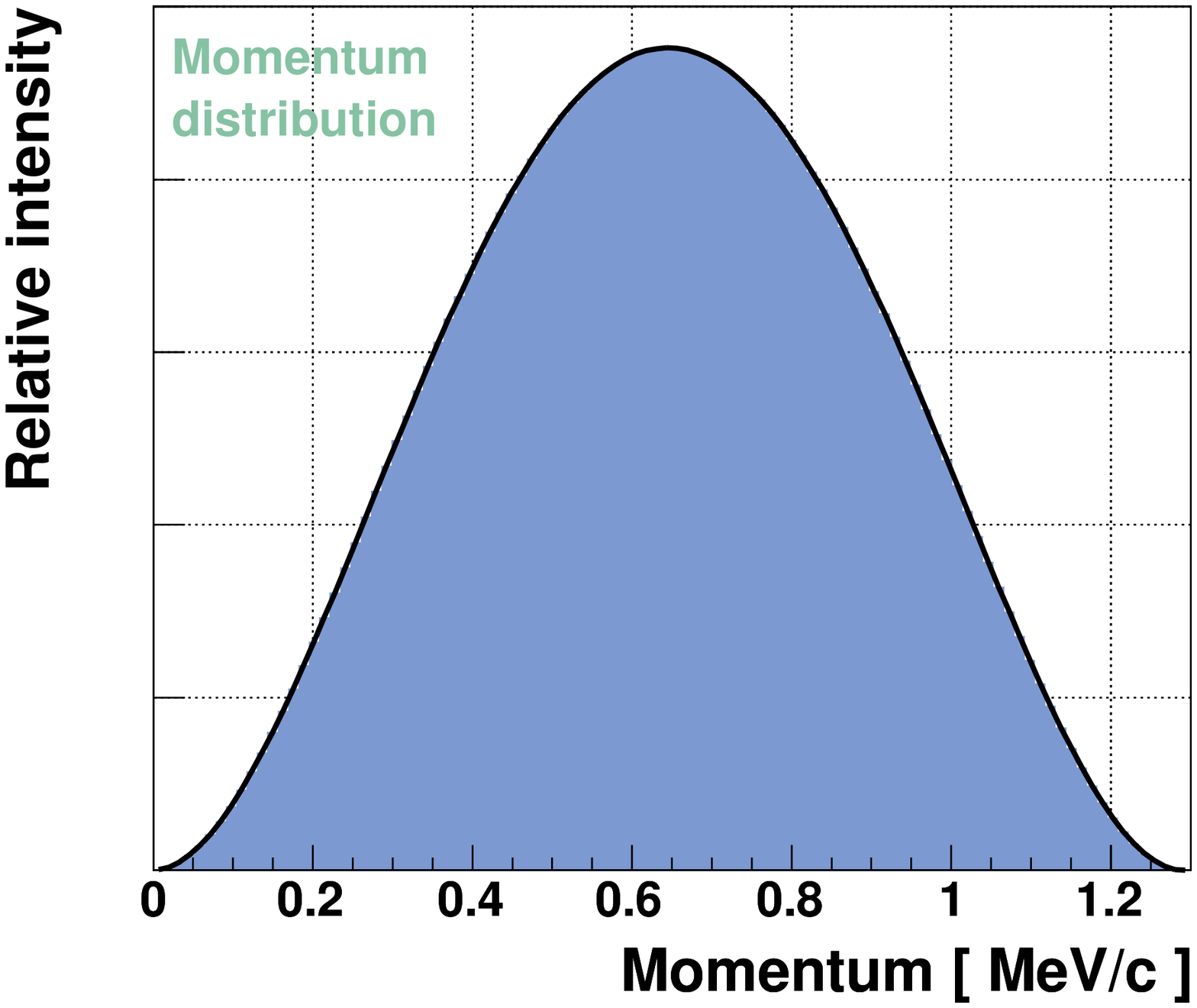}
  \includegraphics[width=0.49\textwidth]{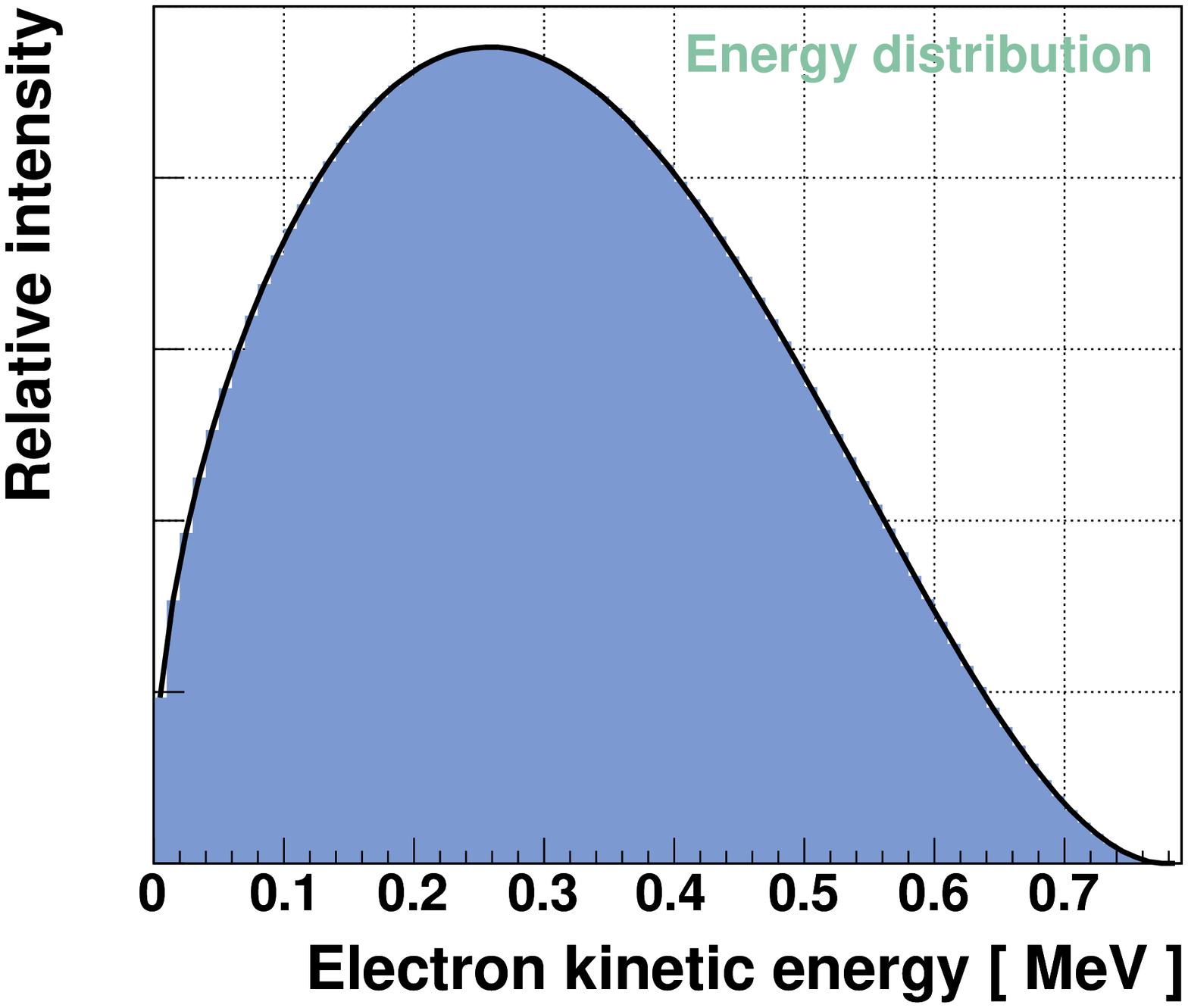}
  \caption{The observed momentum and energy distributions for the $\beta$ electron}
\end{figure}

The neutron is about 0.2\% more massive than a proton, which translates to an energy difference 
of 1.29 \MeV. Therefore, from the energy conservation law, it is possible for a free neutron to 
decay with the emission of an electron and an electron anti-neutrino
\[n\rightarrow p+e^-+\bar{\nu}_e \mbox{\ \ \ \ \ Q = 782.2$\,\pm\, $0.1 \keV}.\] 
Because it is a three-particle decay, produced electrons have continuous distribution of 
momentum and energy (Fig. \ref{fig-beta}) given by
\begin{eqnarray}\label{eq-beta1}
  W_1(p_e)&\propto& F(Z,p_e)p_e^2(p_{max}-p_e)^2,\\
  W_2(E_e) &\propto& \frac{2\pi\alpha}{\beta}\frac{1}{1-e^{-\frac{2\pi\alpha}{\beta}}}p_eE_e(E_{max}-E_e)^2,
\end{eqnarray}
where $ F(Z,p_e)$ is the Fermi function, $\beta=\frac{p_e}{E_e}$, $E_{max}=1.29~\MeV$, $p_{max}=1.19~\MeV/c$ 
and $\alpha$ is the fine structure constant.

\begin{figure}[!h]
  \center
  \includegraphics[scale=0.7]{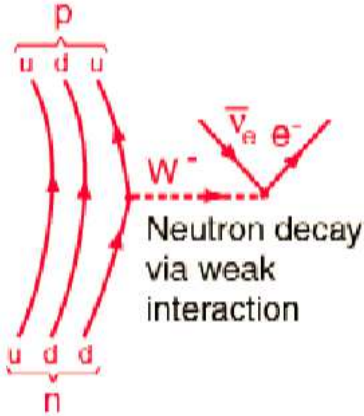}
  \caption{The neutron decay as the transformation of one of the neutron's down quarks into 
    an up quark.\label{fig-feynman}}
\end{figure}

The decay of neutrons involves the weak interaction (see Fig. \ref{fig-feynman}), which, according 
to the theoretical description embedded in the SM (Feynman and Gell-Mann, 1958), has the strict 
vector-axial form ($V-A$).
To find $\cal{T}$-violating effects in $\beta$ decay, one has to measure at least three
vector or axial-vector quantities, namely momenta and spins of the different particles involved
in the decay ($e$, $\nu_e$, $p$, $n$); the quantities, which are all reversed under the time reversal 
operation.
Rotational invariance requires the observables to be scalars or pseudo-scalars, therefore the lowest
order $\cal{T}$-violating combination of spins and momenta appears in the form of the mixed triple product.
From the experimentally accessible quantities, four such products can be formed (see the Tab. \ref{tab-correlations}).
\begin{table}[!h]
\center
\begin{tabular}{|c|c|l|}
\hline\hline
Correlation & Broken symmetry & Definition \\\hline & &\\
$R$ & $\cal{T}$, $\cal{P}$ & $\vec{J}\cdot\left(\vec{p}_e\times\vec{\sigma}\right)$\\& &\\
$D$ & $\cal{T}$ & $\vec{J}\cdot\left(\vec{p}_e\times\vec{p}_{\nu_e}\right)$\\& &\\
$V$ & $\cal{T}$, $\cal{P}$ & $\vec{J}\cdot\left(\vec{P}\times\vec{\sigma}\right)$\\& &\\
$L$ & $\cal{T}$ & $\vec{P}\cdot\left(\vec{p}_e\times\vec{\sigma}\right)$ \\& &\\
\hline\hline
\end{tabular}
\caption{$\cal{T}$-violating triple products. $\vec{J}$ is the neutron spin, $\vec{p}_e$ and $\vec{\sigma}$
are the momentum and spin of the electron, respectively, $\vec{P}$ denotes the proton momentum and $\vec{p}_{\nu_e}$
stands for the momentum of the neutrino. \label{tab-correlations}}
\end{table}

\begin{figure}[h]
  \center
  \includegraphics[scale=0.9]{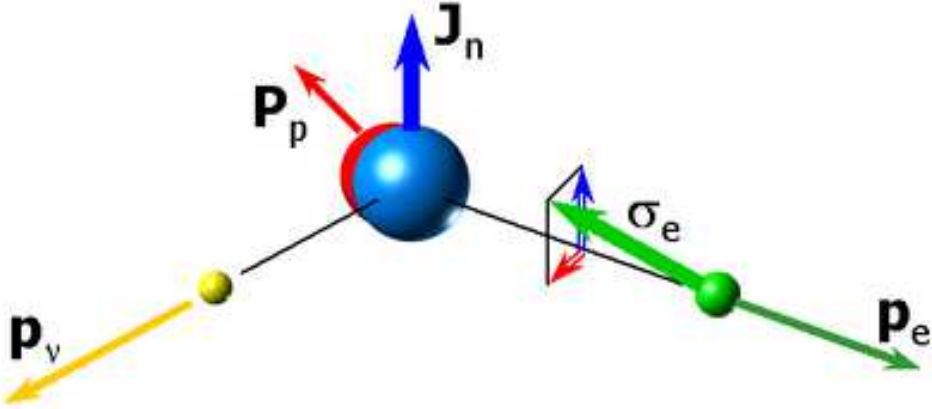}
  \caption{Directions of spins of the neutron and the electron from the $\beta^-$ decay. In the $R$-correlation 
    experiment the transversal component of the electron polarisation will be measured\label{fig-decay}}
\end{figure}

From now on let's concentrate on the planned experiment, in which the momentum and spin of electrons from
decaying oriented neutrons will be measured.
The decay probability distribution suitable for this case has been derived by Jackson \cite{jackson}
\begin{equation}\label{eq-w}
W\propto\left[1+
  b\frac{m}{E_e}+
  A\frac{\vec{\left<J\right>}\cdot\vec{p}_e}{E_e}+
  G\frac{\vec{\sigma}\cdot\vec{p}_e}{E_e}+
  N\vec{\sigma}\cdot\vec{\left<J\right>} +
  Q\frac{\vec{\sigma}\cdot\vec{p}_e}{E_e+m}\frac{\vec{\left<J\right>}\cdot\vec{p}_e}{E_e} +
  R\frac{\vec{\left<J\right>}\cdot\left(\vec{p}_e\times\vec{\sigma}\right)}{E_e}\right]  
\end{equation}
where \\\\
\begin{tabular}{ccll}
  $b$&&&is the Fierz interference coefficient,\\
  $A$&=&$-2\frac{\lambda^2+\lambda}{1+3\lambda^2}$&is the decay asymmetry parameter,\\
  $G$&=&$-1$,&\\
  $Q$&=&$2\frac{1+\lambda}{3+\lambda^2}$,&\\
  $N$&=&$-A\frac{m_e}{E_e}$,&\\
  $\lambda$&&&denotes the coupling constants ratio $C_A/C_V$.\\
\end{tabular}\\\\
Using $\lambda=-1.267$ (see \cite{pdg}), one retrieves values:
\[A=-0.1162, \mbox{\ \ \ \ } Q=-0.1160.\] The coefficient $b$ vanishes if there are no scalar and tensor couplings, 
therefore the zero value has been used.
The formula \ref{eq-w} is the consequence of the interaction Hamiltonian density for weak currents, as given by 
Yang and Lee \cite{lee}.
As can be seen on the Fig. \ref{fig-decay}, $R$ and $N$ parameters are proportional to orthogonal components of
the transversal electron polarisation.
However, only the $R$-correlation reveals direct sensitivity to the existence of exotic $\cal T$-violating scalar 
and tensor couplings in the semileptonic weak interactions.

\section{Polarisation theory}
The following brief review of the polarisation theory formalism is
based on the Refs. \cite{mcmaster}, \cite{stehle} and
\cite{hoogduin}. The Stokes parameters are defined below for the
general case and for the specific case of electrons. When used as
a four-vector, the Stokes parameters allow the ordinary
polarisation-sensitive cross sections to be written in matrix
form, which is a very convenient representation for the
description of polarisation phenomena. 

\subsection{Stokes parameters}
In quantum mechanics the wave function describing a pure state of polarisation can be expanded
in a complete set of orthonormal eigenfunctions. For particles of spin $\frac{1}{2}$ this expansion
contains only two terms,
\begin{equation}
  \psi = a_1 \psi _1 + a_2 \psi_2.
\end{equation}
The wave functions describing pure states may be chosen in the form
\begin{eqnarray}\label{partial-func}
  \psi_1 = \left(\begin{array}{c}1\\0\end{array}\right) &  &
  \psi_2 = \left(\begin{array}{c}0\\1\end{array}\right).
\end{eqnarray}
Thus, the general wave function describing the beam is given by
\begin{equation}\label{gen-func}
  \psi = \left(\begin{array}{c}a_1\\a_2\end{array}\right).
\end{equation}
In the specific case of the electron, the explicit form of this
function can be found by solving the Dirac equation. As a result,
one can achieve two independent solutions (spin ``$+$'' and spin
``$-$'') for the electron and two for the positron. Choosing the
solutions for the electron, one immediately obtains wave functions
(\ref{partial-func}).

In the rest frame of the electron the four-component solutions of
the Dirac equation are reduced to two-component spinors, leading to the
following expressions for the expectation value of the unit matrix
and the expectation values of the Pauli spin operators:
\begin{eqnarray*}
  I = \langle\psi\mid I\mid\psi\rangle =
  \left(\begin{array}{cc}a_1^* & a_2^*\end{array}\right)\pspint\left(\begin{array}{c}a_1\\a_2\end{array}\right) & = & a_1 a_1^* - a_2 a_2^* \,,\\
  P_1 = \langle\psi\mid\sigma_z\mid\psi\rangle =
  \left(\begin{array}{cc}a_1^* & a_2^*\end{array}\right)\pspinz\left(\begin{array}{c}a_1\\a_2\end{array}\right) & = & a_1 a_1^* - a_2 a_2^* \,,\\
  P_2 = \langle\psi\mid\sigma_x\mid\psi\rangle =
  \left(\begin{array}{cc}a_1^* & a_2^*\end{array}\right)\pspinx\left(\begin{array}{c}a_1\\a_2\end{array}\right) & = & a_1 a_2^* + a_2 a_1^* \,,\\
  P_3 = \langle\psi\mid\sigma_y\mid\psi\rangle =
  \left(\begin{array}{cc}a_1^* & a_2^*\end{array}\right)\pspiny\left(\begin{array}{c}a_1\\a_2\end{array}\right) & = & \iu(a_1 a_2^* - a_2 a_1^*).
\end{eqnarray*}
This is the set of four so-called ``Stokes parameters'' which
represent observables and completely characterise the electron
in a pure polarisation state. As it can be seen from the
definition, the physical interpretation of the parameter $I$ is
quite obvious, it is simply the total beam intensity. Moreover, if
the electron is transformed to its rest frame,
$\vec{P}=(P_1,P_2,P_3)$ can be considered as the spin direction.
Note, that the beam in a pure state is completely polarised,
therefore $|\vec{P}|=1$.

\subsection{Mixed states}
A partially polarised beam of electrons cannot be represented by a
single wave function, but by an incoherent ``ensemble'' of pure
states, each characterised by its own wave function. In order to
describe such a case, the {\em density matrix}\/ formalism is used
(for detailed introduction see \cite{tolhoek}). The density matrix
$\rho$ is a $2\times2$ hermitian matrix with positive or zero
eigenvalues and trace equal 1. For the special case of a {\em
totally polarised beam}\/, corresponding to the function
(\ref{gen-func}), the density matrix is
\[\rho=\left(\begin{array}{cc}
  a_1^* a_1 & a_2^* a_1 \\
  a_1^* a_2 & a_2^* a_2\end{array}\right),\]
which can be always brought into the simple form
\[\rho=\left(\begin{array}{cc}
  1 & 0 \\
  0 & 0\end{array}\right),\]
by a unitary transformation. Although, in the most general case,
$\rho$ can be always diagonalised, for mixed states both diagonal
elements remain nonzero. The resulting matrix can be considered as
the incoherent superposition of the unpolarised and totally
polarised beam
\[\rho=\left(\begin{array}{cc}
  \rho_a & 0 \\
  0 & \rho_b\end{array}\right) =
  (1-P)\left(\begin{array}{cc}\frac{1}{2} & 0\\0 &
  \frac{1}{2}\end{array}\right)+
  P\left(\begin{array}{cc}1 & 0\\0 & 0\end{array}\right),\]
where $P$ ($0\leq P=\rho_a-\rho_b\leq 1$) is called the {\em
degree of polarisation}.

The density matrix can be also expressed using the Stokes
parameters
\[\rho = \frac{1}{2}(I+\vec{P}\cdot\vec{\sigma}),\]
it is worth mentioning that now, for mixed states, the
polarisation vector norm $|\vec{P}|\equiv~P<1$.
\subsection{Matrix representation}
\begin{table}
    \begin{tabular}{|c|cl|}
      \hline\hline
      Stokes parameter & & Interpretation \\\hline & &\\
      $I$ & & Intensity\\& &\\
      $P_1$ & $\updownarrow_{-1}^{+1}$ & Longitudinal polarisation (spin in $z$ direction)
      \\& &\\
      $P_2$ & $_{+1\swarrow}\!^{\nearrow-1}$ & Transversal polarisation (spin in $x$ direction)
      \\& &\\
      $P_3$ & $^{-1}\longleftrightarrow^{+1}$ & Transversal polarisation (spin in $y$
      direction) \\& &\\
      \hline\hline
    \end{tabular}
    \caption{Interpretation of components of the Stokes vector.\label{tab-stokes}}
\end{table}
The Stokes parameters are very often written in the form of a
four-vector: \[\left(\begin{array}{c}  I \\P_1 \\P_2 \\P_3
\end{array}\right) \equiv \left(\begin{array}{c}I\\\vec{P} \end{array}\right).\]
The interpretation and some typical examples are given in the table \ref{tab-stokes}
and below:\\\\
$\left(\begin{array}{c}1\\0\\0\\0\end{array}\right)$ represents an
unpolarised beam,\\\\
$\left(\begin{array}{c}1\\\pm1\\0\\0\end{array}\right)$ or
$\left(\begin{array}{c}1\\0\\\pm1\\0\end{array}\right)$ describe
transversal polarisation in $z$ and $x$ directions.

Because the Stokes parameters are dependent on the polarisation basis one has chosen, there 
exists a transformation matrix $M$, which relates a Stokes vector in one reference frame to the 
same one in another coordinate system:
\[\left(\begin{array}{c}I'\\\vec{P}' \end{array}\right) = M\left(\begin{array}{c}I\\\vec{P} \end{array}\right).\]
For instance, if the second coordinate system is rotated about the direction of
propagation at an angle $\theta$ to the right, then the matrix M is a simple rotation matrix 
given by
\begin{equation}\label{eq-rotation}
  M = \left(\begin{array}{cccc}1&0&0&0\\0&1&0&0\\0&0&\cos\theta&\sin\theta\\0&0&-\sin\theta&\cos\theta\end{array}\right).
\end{equation}
Without any difficulties one can write similar matrices for rotations about remaining axis.

The biggest advantage of the Stokes vector formalism is the possibility of calculating 
probabilities and cross-sections in a convenient and intuitive way, using scalar products and
simple matrix operations. The formula
\[W = \frac{1}{2}\left(\begin{array}{cc}1&\vec{D}\end{array}\right)\left(\begin{array}{c}I\\\vec{P}\end{array}\right),\]
gives the probability of detecting a particle characterised by the Stokes parameters 
$(1$~$\vec{D})$ in a beam characterised by parameters $(1$~$\vec{P})$. In other words, the vector
$(1$~$\vec{D})$ describes properties of a detector. Of course, a polarisation-insensitive 
detector corresponds to the Stokes vector $(1~0~0~0)$; in this case one effectively measures 
each of two orthogonal states and adds the probabilities, leading to the formula:
\[W = \left(\begin{array}{cccc}1&0&0&0\end{array}\right)\left(\begin{array}{c}I\\\vec{P}\end{array}\right).\]

Finally, an interaction can be introduced here in a very natural way. When a particles 
experience a polarisation-sensitive interaction, then, in general, their Stokes vector is 
transformed by a $4\times$4 matrix $T$, which depends on the interaction type
\begin{equation}\label{eq-depol}
  \left(\begin{array}{c}I\\\vec{P} \end{array}\right) = T\left(\begin{array}{c}I_0\\\vec{P_0} \end{array}\right).
\end{equation}
Thus, the probability of detecting the beam in a state $(1$~$D)$ after the interaction $T$ is 
given by: 
\[W = \frac{1}{2}\left(\begin{array}{cc}1&\vec{D}\end{array}\right)T\left(\begin{array}{c}I\\\vec{P}\end{array}\right).\]
Full interaction matrices for all processes specific to electrons, positrons and photons are 
available in the Ref. \cite{mcmaster}. However, later on in this thesis, {\em reduced}\/ 
matrices (given in \cite{hoogduin}) will be presented and used. The upper left element of a 
reduced $T$ matrix is always equal to one or, precisely speaking, the cross section of an 
unpolarised beam detected by a polarisation-insensitive detector is normalised to unity. 
\\\\
To summarise the main ideas of this section, we note that:
\begin{itemize}
  \item the Stokes parameters describe the polarisation of the beam in a unified way,
  \item the interactions can be introduced to the formalism as matrices,
  \item $(I~\vec{P})$ can be calculated from the positive-energy components of the 
    Dirac equation solution,
  \item an unpolarised beam can be considered as an ensemble of electrons with spin $\vec{P}$
    pointing isotropically in all directions.
\end{itemize}

\section{Passage of electrons through matter}\label{sec-passage}
As it will be described in the section \ref{sect-experiment}, the energy, momentum
and polarisation of the electron from $\beta^-$ decay are essential for the $R$-correlation 
determination.
However, before the electron is detected, it has to cross the whole experimental apparatus and in
the meantime undergoes numerous interactions with matter. Of course, electron energy, momentum 
and polarisation could have been changed due to these processes and hence one has 
to take them into account in order to retrieve the primary electron properties. 
In the energy region which is in our concern (below 780~\keV), the dominant processes are 
ionisation (M{\o}ller scattering) and multiple elastic Coulomb scattering. Only a few
percent of electrons lose their energy by bremsstrahlung.

Physics of all these processes is well known and has been already
implemented in the \geant, except the depolarisation phenomena. The following section contains some 
fundamental formulae, which the package is based on (for more details see \cite{geant-physics}). 
Additional polarisation effects have been incorporated in the simulation part responsible for 
the electron transportation by means of depolarisation matrices, which are given in 
\cite{hoogduin} and explicitly shown below.

Last but not least, polarised $\beta^-$ decay electrons are Mott-scattered in the analysing lead 
foil, which is actually the main idea of the measurement. Therefore, the last section covers
the topic of the spin dependent electron scattering on nuclei and its analysing power.

\subsection{M{\o}ller scattering}
This process occurs when an incident electron is inelastically scattered on an atomic 
electron from the target material. The value of the maximum energy transferable to a free electron\footnote{Note that in the M{\o}ller process the scattering and scattered electrons are 
  indistinguishable. However, the highest-energy member of the scattered particles is arbitrarily
  associated with the original incoming electron.} 
equals $(E_{incoming} - m_ec^2)/2$. If the transfered energy is much larger than the excitation
energy of the material, the atom is ionised and the electron is emitted as a so called 
``$\delta$-ray''. The total cross section per atom for M{\o}ller scattering is given by
\[\sigma(Z,E,T_{cut})=\frac{2\pi r^2_eZ}{\beta^2(\gamma-1)}\left[\frac{(\gamma-1)^2}{\gamma^2}\left(\frac{1}{2}-x\right)+\frac{1}{x}-\frac{1}{1-x}-\frac{2\gamma-1}{\gamma^2}\ln\frac{1-x}{x}\right], \]
where:
\begin{center}
  \begin{tabular}{cclccl}
    $\gamma$ & = & $E/mc^2,$ & 
    $\beta^2$ & = & $1 - (1/\gamma^2)$,\\
    $x$ & = & $T_{cut}/(E-mc^2)$,&
    $y$ & = & $1/(\gamma+1)$,\\
    $r_e$ & = & the classical radius of the electron.& & &
  \end{tabular}
\end{center}
$T_{cut}=1$~\keV\ is a threshold kinetic energy, below which the process is considered as a 
continuous (in such a case different formulae are used and $\delta$-rays are not simulated).

The differential cross section,
\[ \frac{d\sigma}{d\epsilon}=\frac{2\pi r^2_eZ}{\beta^2(\gamma-1)}\left[ \frac{(\gamma-1)^2}{\gamma^2}+\frac{1}{\epsilon}\left(\frac{1}{\epsilon}-\frac{2\gamma-1}{\gamma^2}\right)+\frac{1}{1-\epsilon}\left(\frac{1}{1-\epsilon}-\frac{2\gamma-1}{\gamma^2}\right)\right]\]
where $\epsilon_0=\frac{T_{cut}}{E-mc^2}\leq\epsilon\leq\frac{1}{2}$, is used to sample the 
$\delta$-ray energy and direction. With respect to the direction of the incoming particle, 
the $\delta$-ray azimuthal angle $\phi$ is generated isotropically and the polar angle 
$\theta$ is calculated from energy-momentum conservation, which corresponds to a situation
where target electrons are unpolarised. That is how energy and momentum of both, incident
and ejected particle is calculated.

Given these values, one can calculate the depolarisation of the incident electron using 
the formula \ref{eq-depol} and the reduced depolarisation matrix $T_{M}$\\
\[
T_M = \left(\begin{array}{cccc}
  1 & 0 & 0 & 0 \\
  0 & C & D & 0 \\
  0 & D & E & 0 \\
  0 & 0 & 0 & F
\end{array}\right),\]
where:
\begin{center}
  \begin{tabular}{ccl}
    $C$ & = &$ 2\cos\theta(2\gamma^2 -1)(2\gamma^2 -1-\gamma^2 \sin^2\theta)/I,$\\
    $D$ & = &$ 2\gamma(2\gamma^2 -1)\sin\theta\cos^2 \theta/I,$\\
    $E$ & = &$ 2\cos\theta(2\gamma^2 -1)(2\gamma^2 -1-\sin^2\theta)/I,$\\
    $F$ & = &$ 2[(2\gamma^2 -1)^2 -(2\gamma^4 -1)\sin^2 \theta]/I,$\\
    $I$ & = &$ \frac{1}{2}[(2\gamma^2 -1)^2 (4-3\sin^2 \theta)+(\gamma^2 -1)^2 (\sin^4 \theta+4\sin^2 \theta)]$
  \end{tabular}
\end{center}
and where $\gamma$ is the electron energy (units of $mc^2$) and $\theta$ is the scattering 
angle, both in the centre-of-mass frame (CM). The relations between the CM (not primed) and laboratory 
(primed) quantities $\gamma'$ and $\theta'$ are:
\[\gamma' = 2\gamma^2-1,\]
\[\cos\theta = \frac{2-(\gamma'+3)\sin^2 \theta'}{2+(\gamma'-1)\sin^2\theta'}.\]
One can show that the following relation between the corresponding solid angles holds:
\[ d\Omega = \frac{8(\gamma'+1)\cos\theta'}{[2+(\gamma'-1)\sin^2 \theta']^2}d\Omega'.\]

The Stokes vector of the incoming particle should be transformed to the coordinate system,
where the direction of movement is along the $z$-axis and the $xz$-plane is the plane of the 
scattering. The polarisation of the final electron is given in the new reference frame
rotated through the scattering angle $\theta$ about the $y$ axis.

\subsection{Multiple scattering}
In addition to inelastic collisions with the atomic electrons, charged particles passing 
through matter also suffer from repeated elastic Coulomb scatterings on nuclei although
with a somewhat smaller probability. Each of them is individually governed by the well-known
Rutherford formula (when ignoring, for simplicity, spin effects and screening)
\begin{equation}\label{eq-rutherford}
  \frac{d\sigma_{\mbox{\tiny{\it R}}}}{d\Omega}(\theta)=
  \frac{1}{4\pi\varepsilon_0}\left(\frac{Z_1Z_2e^2}{4E_{kin}}\right)^2\frac{1}{\sin^4\frac{\theta}{2}}.
\end{equation}
Moreover, since
the nucleus is much more massive than the scattered electron, the energy transfer in the process
is negligible. 

If the average number of independent scatterings is greater than 20, the problem can be treated
statistically to obtain a probability distribution for the angle of deflection as a function
of the particle step length in the traversed material. 
The method governing the multiple scattering (MSC) of charged particles in matter used in 
\geant\ is based on Lewis' theory (see \cite{lewis}) and uses model functions to determine the 
angular and spatial distributions after each step of tracking. The model functions have been 
chosen in such a way as to give the same moments of the distributions as the Lewis theory.

The properties of the MSC process are completely determined by the transport mean free paths
$\lambda_k$, which are functions of the energy in a given material. The $k$-th transport mean
free path is defined as
\[\frac{1}{\lambda_k} = 2\pi n_a \int_{-1}^{1}[1-P_k(\cos\chi)]\frac{d\sigma(\chi)}{d\Omega}d(\cos\chi) \]
where $d\sigma(\chi)/d\Omega$ is the differential cross section of the single scattering (e.g. 
the Rutherford formula), $P_k(\cos\chi)$ is the $k$-th Legendre polynomial and $n_a$ is the 
number of scattering centres per volume. The mean value of the {\em geometrical path length}
\/\footnote{The shortest, straight line distance between the endpoints of a single step.} 
corresponding to a given {\em true path length}
\/\footnote{The path length of an actual particle usually longer than the geometrical path 
length, since the path is random and zigzag.} 
is given by
\[ \left<z\right> = \lambda_1 \left[1-\exp\left(-\frac{t}{\lambda_1}\right)\right].\]
At the end of the true step length $t$, the scattering angle is $\theta$, the mean value and
variation of its cosine are
\[ \left<cos\theta\right> = \exp\left(-\frac{t}{\lambda_1}\right)\]
\[\sigma^2 = \left<\cos^2\theta\right> - \left<\cos\theta\right>^2 = \frac{1+2e^{-2\kappa\tau}}{3}-e^{-2\tau},\]
where $\tau=t/\lambda_1$ and $\kappa=t/\lambda_2$.
In addition to this, the square of the mean lateral displacement (assuming the the initial
momentum is parallel to the $z$ axis) is 
\[ \left<x^2+y^2\right> = \frac{4\lambda_1^2}{3}\left[\tau-\frac{\kappa+1}{\kappa}+\frac{\kappa}{\kappa-1}
e^{-\tau}-\frac{1}{\kappa(\kappa-1)}e^{-\kappa\tau}\right].\]

The angular distribution of the scattered particle is sampled according to a model function
$g(u)$, where $u=\cos\theta$. The functional form of $g$ is
\[ g(u)=p[qg_1(u)+(1-q)g_3(u)]+(1-p)g_2(u),\]
where $0\leq p,q\leq 1$,  $g_i$ are functions of $u$, normalised over the range 
\mbox{$u\in[-1,1]$}
\begin{center}
\begin{tabular}{ccc}
$g_1=C_1e^{-a(1-u)}$, & $g_2=C_2\frac{1}{(b-u)^c}$, & $g_3=C_3$,\\
\end{tabular}
\end{center}
$a,b,c>0$ and $C_i$ are normalisation constants. For more details see 
the Ref. \cite[section 6.2]{geant-physics}, regarding the purposes of this thesis, it is 
enough to mention that for small scattering angles $g_1(u)$ is nearly Gaussian, while for 
large angles $g_2(u)$ has a Rutherford-like tail.

Finally, the polarisation change in the MSC process is calculated in a very simple way 
(see \cite[section 4.3.1]{hoogduin}). When the momentum vector of the particle is rotated
over an angle $\phi$, its component of the polarisation vector in the scattering plane is rotated
over an angle $\theta$ given by 
\[\theta = \phi\frac{E_{e} - 1}{E_{e}}, \]
where $E_{e}$ is the total energy of the particle in units $mc^2$. A simple observation 
indicates that in the higher energy limit, the polarisation of a longitudinally polarised beam 
follows its momentum, while for lower energies, the slower rotation of the polarisation vector 
becomes significant.

\subsection{Bremsstrahlung}
For $\beta^-$ decay electrons, the radiation of photons in the field of nucleus is much less 
probable process than ionisation or the multiple scattering and affects only few percent of 
electrons, causing minor energy loses and depolarisation. Below 1 \MeV\ the theoretical description
of bremsstrahlung has to be considered as an approximation, the particular model 
exploited in \geant\ \cite[section 7.2]{geant-physics}  results in up to 15\% errors for both 
the cross section and the energy loss. It is based on the tabulated cross sections of Seltzer
and Berger (Ref. \cite{seltzer}), together with the Bethe-Heitler formula (which includes
the dielectric suppression of the radiation) and the correction for the Landau-Pomeranchuk-Migdal
effect\footnote{The suppression of photon production due to the interference of radiation emitted
before and after the multiple Coulomb scattering event.}. The angular distribution of the 
emitted photon momentum was reported by Tsai \cite{tsai} and its simplified version is 
implemented in \geant.
 
The depolarisation matrix specific to bremsstrahlung was derived in the Ref. 
\cite[section 3.9.2]{hoogduin} and is taken as
\[T_{brem, e} = \left(\begin{array}{cccc}
  1 & 0 & 0 & 0 \\
  0 & G+H & F & 0 \\
  0 & E & G & 0\\
  0 & 0 & 0 & G
\end{array}\right) \]
where:
\begin{center}
  \begin{tabular}{ccl}
    $I$ &=& $(\epsilon_1 ^2+\epsilon_2 ^2)(3+2\Gamma)-2\epsilon_1 \epsilon_2(1+4u^2 \xi^2 \Gamma)$,\\
    $E$ &=& $4k\xi\Gamma\epsilon_1 u(2\xi-1)/I$,\\
    $F$ &=& $4k\xi\Gamma\epsilon_2 u(1-2\xi)/I$,\\
    $G$ &=& $4\epsilon_1 \epsilon_2 [(1+\Gamma)-2u^2 \xi^2 \Gamma]/I$,\\
    $H$ &=& $k^2 [1+8\Gamma(\xi-\frac{1}{2})^2 ]/I$
  \end{tabular}
\end{center}
and:
\begin{center}
  \begin{tabular}{ccl}
    $\epsilon_1$ & = & total energy of the incoming electron (in units $mc^2$), \\
    $\epsilon_2$ & = & total energy of the outgoing electron (in units $mc^2$), \\
    $\vec{p}$    & = & electron initial momentum (in units $mc$), \\
    $\vec{k}$    & = & photon momentum (in units $mc$), \\
    $\vec{u}$    & = & component of $\vec{p}$ perpendicular to $\vec{k}$,\\
    $u$          & = & $|\vec{u}|$,\\
    $k$          & = & $\epsilon_1-\epsilon_2$, photon energy (in units $mc^2$), \\
    $\xi$        & = & $\frac{1}{1+u^2}$.
  \end{tabular}
\end{center}
Both incoming and outgoing polarisation vectors are rotated here into the frame defined by
the scattering plane ($xz$) and the direction of the outgoing photon ($z$-axis). Described
approach to the depolarisation by bremsstrahlung contains the Coulomb and screening effects, 
introduced in functions:
\begin{center}
  \begin{tabular}{ccl}
    $\Gamma$ &=& $\ln\frac{1}{\delta}-2-f(Z)+\cal{F}(\delta/\xi)$,\\
    $\delta$ &=& $\frac{k}{2\epsilon_1 \epsilon_2}$,\\
    $f(Z)$ &=& $a^2 [(1+a^2 )^{-1} +0.20206 - 0.0369a^2 +0.0083a^4 -0.002a^6 ]$, 
  \end{tabular}
\end{center}
$f(Z)$, where $a=\alpha Z$, is an approximated form of the Coulomb correction term (see 
\cite{coulomb}). The function $\cal{F}(\delta/\xi)$ is tabularised in literature and 
includes screening effects.
\subsection{Mott scattering}\label{sec-mott}
At relativistic energies, the Rutherford cross section (\ref{eq-rutherford}) describing
the elastic Coulomb scattering is modified by spin effects. The resulting {\em Mott cross 
section}\/ for the electron, derived from the Dirac equation, may be written as
\begin{equation}\label{eq-mott}
  I(\theta)\equiv\frac{d\sigma_{\mbox{\tiny{\it M}}}}{d\Omega}(\theta)=
  \frac{d\sigma_{\mbox{\tiny{\it R}}}}{d\Omega}(\theta)\cdot
  \left(1-\beta^2\sin^2\frac{\theta}{2}\right),
\end{equation}
however, it gives only the spin-averaged cross section. In order to investigate how the 
{\em particularly}\/ polarised electron scatters on a nuclei, one has to introduce the
complex scattering amplitudes $f$ and $g$, satisfying the condition
\[I(\theta)=|f|^2+|g|^2.\] Then, {\em the Sherman function} or so-called {\em analysing power}\/
$S(\theta)$ and the spin rotation functions $T(\theta)$ and $U(\theta)$ are defined as follows:
\begin{center}
  \begin{tabular}{lcr}
    $S(\theta)= \iu\frac{fg^*-f^*g}{|f|^2+|g|^2}$, &
    $T(\theta)= \frac{|f|^2-|g|^2}{|f|^2+|g|^2}$,  &
    $U(\theta)= \frac{fg^*+f^*g}{|f|^2+|g|^2}$.
  \end{tabular}
\end{center}
As one can see, all observable quantities pertaining to the scattering process can be expressed
in quadratic terms of $f$ and $g$. $I$ is the differential cross section for an unpolarised beam,
$S$ is the asymmetry function which gives the transverse polarisation of the scattered electron 
for an unpolarised beam (or the left-right asymmetry for a 100\% transversely polarised beam), 
$T$ and $U$, finally, describe the rotation of the polarisation vector $\vec{P}$ during the 
interaction.
\begin{figure}[h]
  \center
  \includegraphics[scale=0.5]{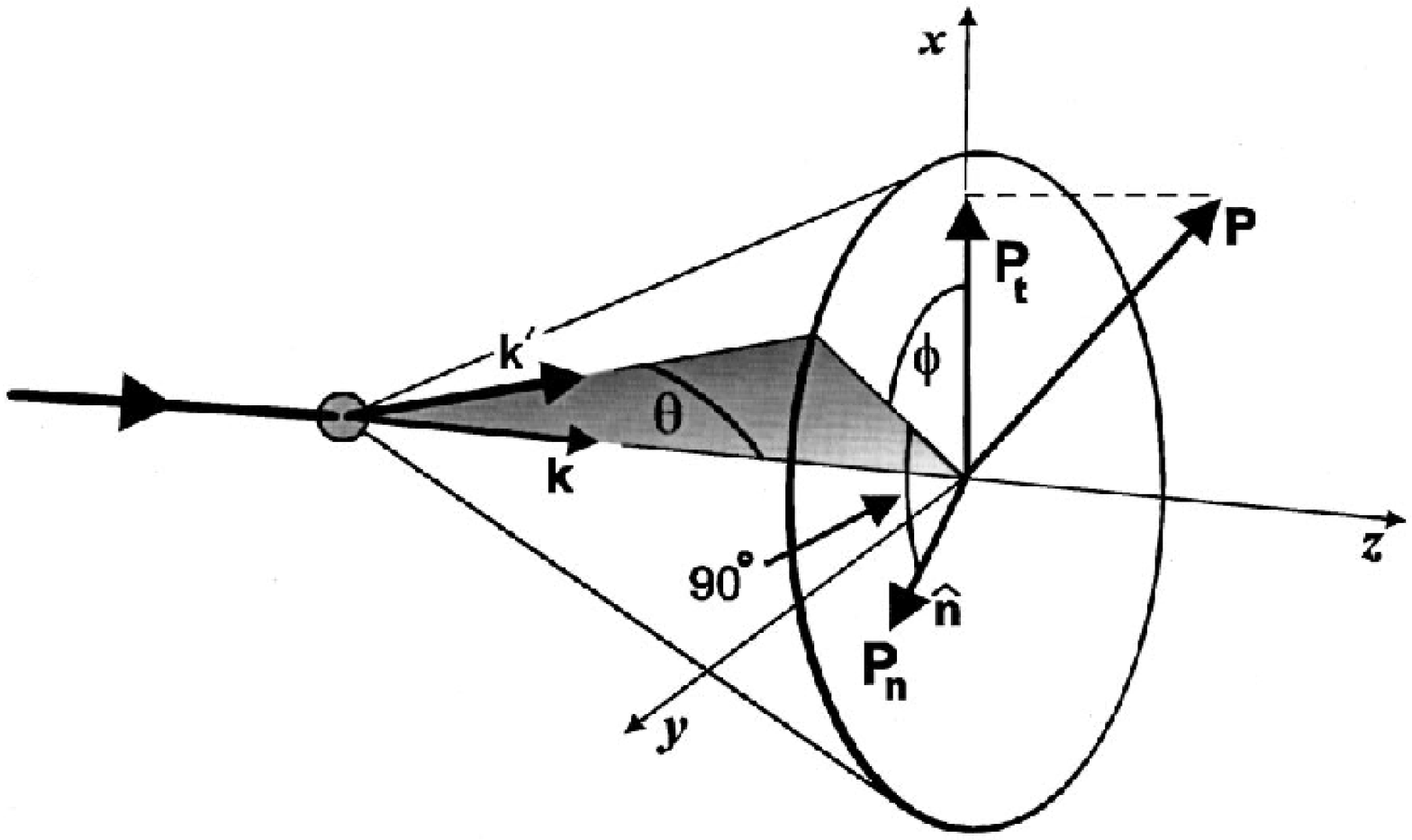}
  \caption{Vectors and angles in Mott scattering (from the Ref.\cite{khakoo}).\label{fig-mott}}
\end{figure}

The differential cross section for the polarised electron is dependent not only on the angle 
of deflection $\theta$ but also on the azimuthal angle $\phi$
\begin{equation}\label{eq-sherman}
\frac{d\sigma}{d\Omega}(\theta,\phi)=
I(\theta)\left(1+S\left(\theta\right)\vec{P}\cdot\hat{n}\right)=
I(\theta)\left(1+S\left(\theta\right)P_n\sin\phi\right),
\end{equation}
where $\hat{n}=\frac{\hat{n_1}\times\hat{n_2}}{\sin\theta}$ is the unit vector normal to the 
scattering plane, $\hat{n_1}$ and $\hat{n_2}$ are unit vectors in the direction of the
electron motion, respectively, before and after scattering and $P_n$ is a component of 
$\vec{P}$ along $\hat{n}$.
After the scattering, the electron polarisation is given by
\begin{equation}
  \vec{P}'=\frac{\left[P_n+S\left(\theta\right)\right]\hat{n}+
  T(\theta)\hat{n}\times[\vec{P}\times\hat{n}]+
  U(\theta)[\hat{n}\times\vec{P}]}{1+\vec{P}\cdot\hat{n}}.
\end{equation}
The $S$, $T$ and $U$ functions has been calculated and tabularised for a variety of elements and 
energies by Sherman \cite{sherman}.
\cleardoublepage

\chapter{Experiment}\label{sect-experiment}
\begin{figure}[h]
  \center
  \includegraphics[width=\textwidth]{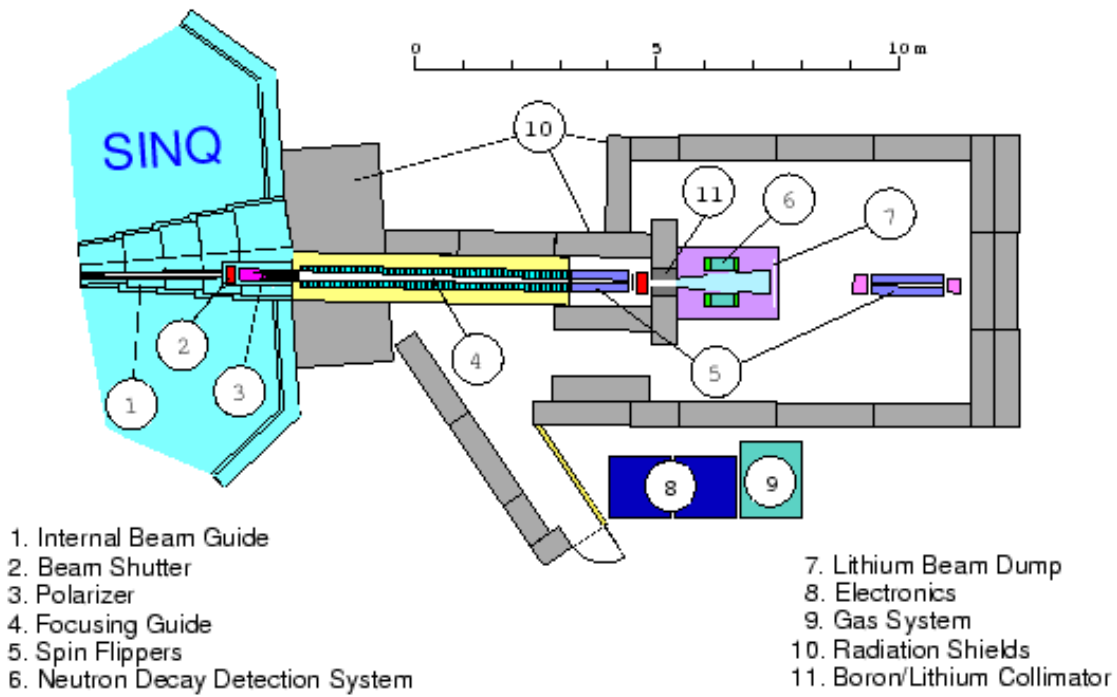}
  \caption{The spallation source SINQ with the polarised cold neutron beam line.\label{fig-sinq}}
\end{figure}

The $R$-correlation measurement is being performed at the Paul Scherrer Institute in Villigen
on the polarised cold neutron beam line of the spallation source SINQ~\cite{ntrv-nn}. Fig. \ref{fig-sinq}
shows the layout of the line and demonstrates where both the polarisation and focusing of the beam 
take place. 

\section{The main principle}
According to section \ref{sec-ntheory}, a measurement of the electron polarisation component, which is 
perpendicular to the plane spanned by the spin of the decaying neutron and the electron momentum, will 
provide an estimation of the $\beta$ decay  $R$-correlation parameter and, if the value was different 
than zero, would discover time reversal symmetry violating processes.
\begin{figure}[h]
  \center
  \includegraphics[width=\textwidth]{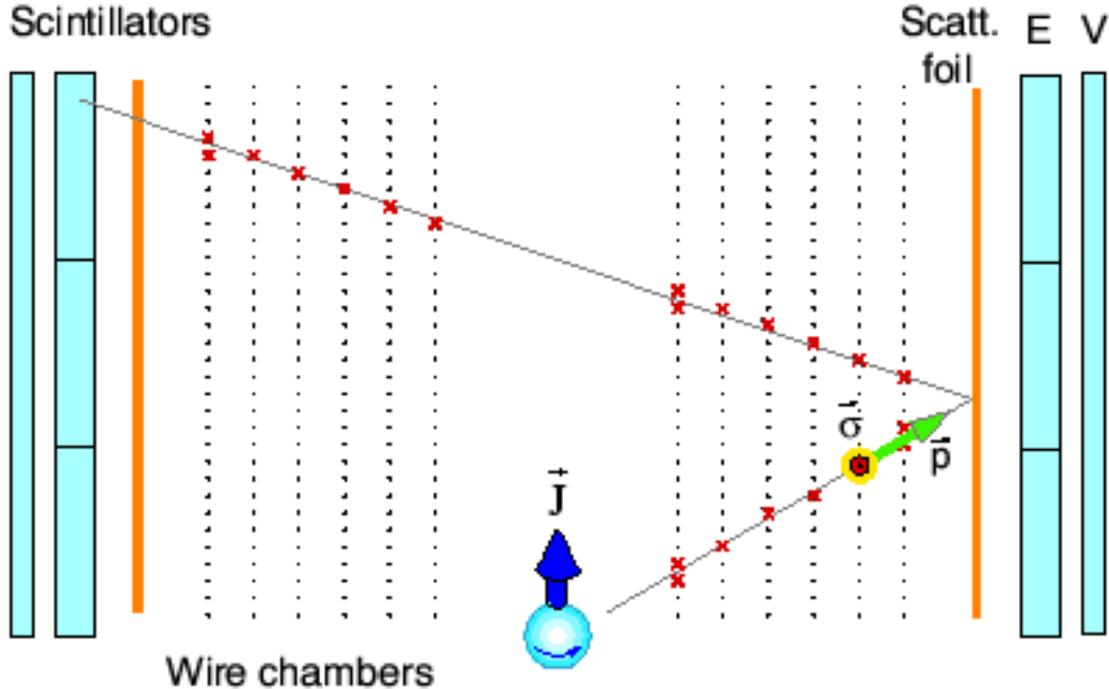}
  \caption{The principle of the measurement. The picture shows so called {\em V-Track}, which
    is the type of event desired for the transversal polarisation component determination.\label{fig-principle}}
\end{figure}
The main principle of the measurement is shown in Fig. \ref{fig-principle}. An electron from the decay 
of a neutron with specified polarisation traverse the wire chamber detector. Afterwards, it scatters
on the analysing foil covered with a layer of lead. Note, that the angle of scattering depends on the 
electron polarisation (Eq. \ref{eq-sherman}). Deflected electrons traverse back the whole setup,
including both wire chambers, the second Pb foil and, finally, hit the scintillator, where the 
remaining energy is left. Electron tracks from wire chambers provide possibility of reconstruction the
emission and Mott scattering angles, which, together with the known neutron beam polarisation, is sufficient 
to obtain the $R$ parameter. In addition, the energy measurement from scintillators serves for 
distinguishing the signal from background, generated mostly by high energy electrons coming from neutron 
captures in surrounding. 

The measurements are done for both neutron spin orientations (``up'' and ``down''),
which helps to cancel out some systematic effects related to the setup geometry and to control better 
the remaining ones. Moreover, another important feature of this experimental configuration is to be used for 
the same purpose. It is clear that from the scattering angles one can acquire 
information on {\em both} transversal components of the electron polarisation, because for both of them 
the experimental technique is exactly the same and both can be measured simultaneously. Thus, 
the $N$-correlation parameter can be extracted from the data, as well as the $R$ parameter. Since the value of $N$
is well known (see Eq. \ref{eq-w}), the comparison with the measured value will serve as an important
clue to systematic errors limiting the experimental accuracy.
Therefore, the behaviour of both electron polarisation components will be studied in the simulation. 

\section{Experimental setup}
The collimated beam of cold neutrons enters the decay chamber filled with helium and covered with a special
material enriched with $^6$Li. Both these elements were chosen to reduce the background, because helium has 
zero cross section for neutron capture and is easy to use and $^6$Li, which  in contrary absorbs neutrons very well, 
does not in effect emit secondary $\gamma$ rays, which could convert to electrons in detector materials. 
In the lithium cover, 
\begin{figure}[h]
  \center
  \includegraphics[width=0.66\textwidth]{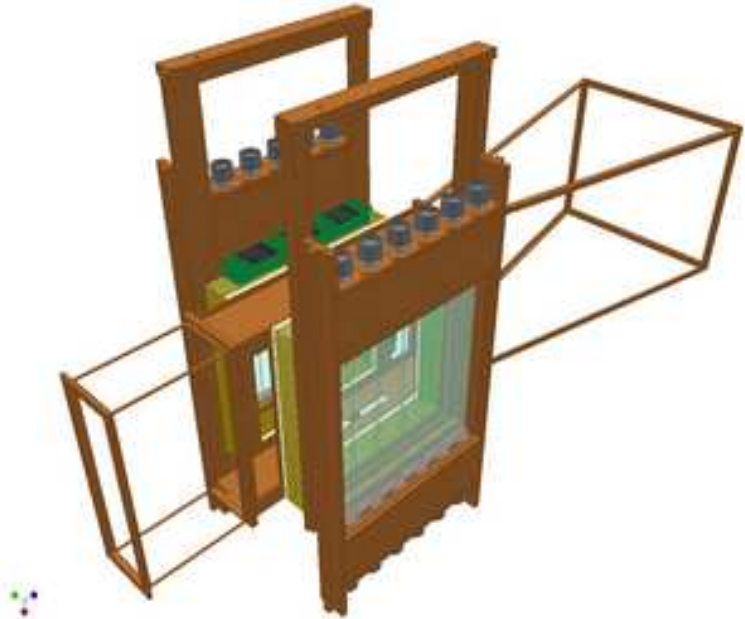}
  \caption{The helium box with MWPC detectors and hodoscopes.\label{fig-setup}}
\end{figure}
on the sides of the decay chamber there are only two mylar windows separating MWPC ({\em MultiWire Proportional Chamber\/}) 
detectors from decay volume. The MWPC chambers are filled with a gas mixture of helium, methylal and isobutan, in proportion optimised
for the best efficiency and stable operation. The analysing Pb foil is placed behind the next mylar window, in a small helium filled box.
In this case, helium is meant to protect the foil from oxidising and is optimal with respect to energy loss ans multiple scattering. 
The last elements of the setup are two systems of position
sensitive scintillators.
\cleardoublepage

\chapter{Simulation}
The comprehensive Monte Carlo simulation of the $R$-correlation experiment has been created by our group since the year 2001 and 
has already been useful for various purposes. Its present version is the sum of individual efforts made by few developers and 
contains all functionality from few recent versions like:
\begin{itemize}
\item {\em EnergyLoss} program for energy calibration, 
\item {\em Sim} used for the investigation of neutron background.
\end{itemize}
The first release of the code and its main part, including extended \geant\ classes for neutron capture, has been written by 
J.~Pulut. E.~Stephan has implemented the first version of the Mott scattering subroutine and the code for generation of the artificial 
wire chamber response.

So far the simulation has been useful for the energy calibration purposes and for the optimisation of the experimental setup
geometry and materials. Now, with the use of new polarised dependent parts, it can be employed for the analysis of systematic effects.
This chapter describes selected features of the simulation with the stress laid on the polarisation transport.

\section{The \geant\ package}
\geant\ (GEometry ANd Tracking) is a widely used set of C++ libraries for simulating the passage of particles through matter, 
originating in the old {\sc Fortran} based {\sc Geant3} version. It includes a complete range of functionality including tracking, 
geometry and physics models and provides tools necessary for implementing a whole experiment in a program. With the use of 
Monte Carlo techniques \geant\ step by step calculates a particle energy losses, direction changes, creates secondary particles and 
simulates physical interactions. In general, the total cross section for the various relevant processes is used to select the process 
that will take place and when it is done, appropriate differential cross sections are used to calculate the kinematics.
However, no electron polarisation sensitivity is included in the code.
The details and further references can be found in the recent paper of the \geant\ project \cite{geant-paper} and on the homepage 
\cite{geant-home}.

\section{Geometry}
\begin{figure}
  \center
  \includegraphics[width=\textwidth, height=0.8\textwidth]{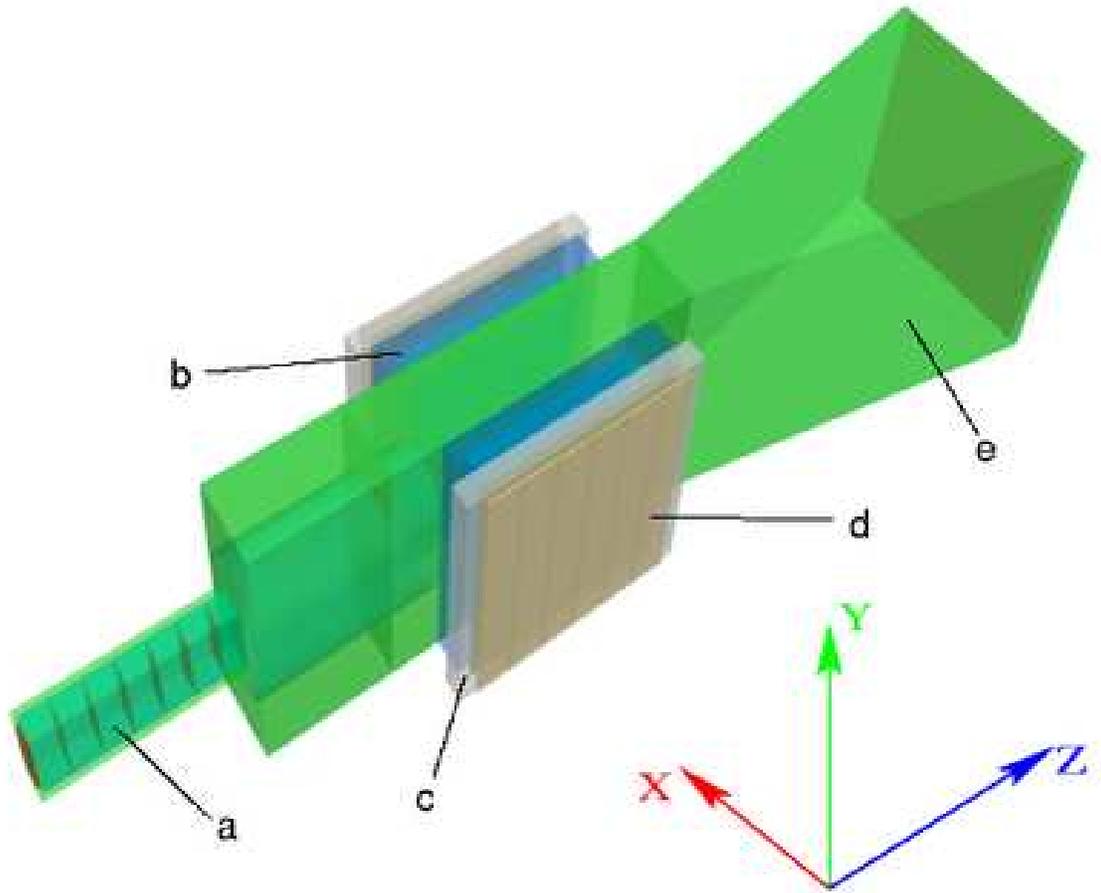}
  \caption{The geometry implemented in the simulation: a) the collimator, b) the proportional chamber,
    c) the helium box with the analysing lead foil, d) scintillators, e) the decay chamber. Directions of axes of the main
    reference system are shown as well. Please note that its origin is located in the middle of the decay chamber, between 
    the wire chambers.
    \label{fig-simgeom}}
\end{figure}

The whole geometry of the experimental apparatus and the neutron beam implemented in the simulation is based on its previous
version (comprehensively described in Ref. \cite{kuzniak}), updated for the new setup. It has been written in a flexible way,
therefore, if needed, the user can easily modify materials and sizes of chosen setup elements or change the distances between them.
All the parameters are available in the macro file {\tt geom.g4mac} presented in appendices and in the source files 
{\tt src/GeometryConstans.icc} and {\tt src/DetectorMaterials.icc} . Of course, modification of the source files 
requires subsequent recompilation of the program. 
Fig. \ref{fig-simgeom} demonstrates the implemented geometry and its details.

\section{Polarisation transport}\label{sec-spintrans}
\begin{figure}
  \center
  \includegraphics[width=0.6\textwidth]{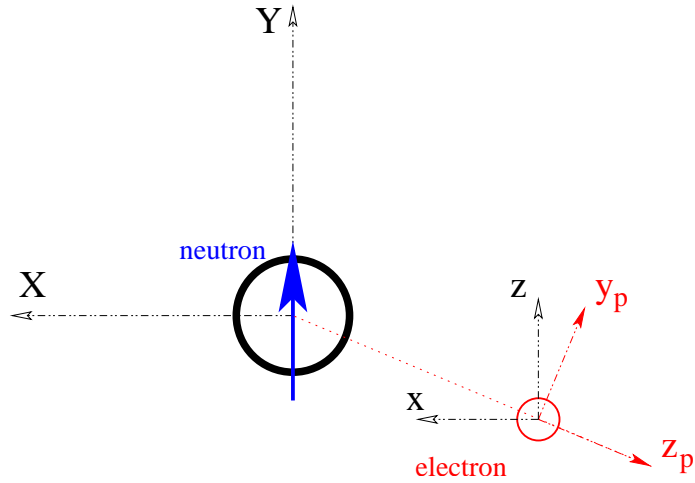}
  \caption{The main reference frame $\{X,Y,Z\}$ and both frames moving with the electron. $\{x_p,y_p,z_p\}$ denotes
    {\em the particle frame}.\label{fig-frames}}
\end{figure}
The electron polarisation $\vec{\sigma}$, sampled as described in section \ref{sec-generation}, is given relative to a right-handed
frame moving with the particle. However, the bookkeeping of all polarisation changes of a particular
electron is performed in a coordinate system with the $z$ axis in the direction of motion, in the frame 
called {\em the particle basis}. Both frames are linked to each other through a simple rotation, given by
a matrix similar to that of Eq. \ref{eq-rotation} (see Fig.\ref{fig-frames}). The full definition of the particle basis 
$\{x_p,y_p,z_p\}$ and its relation to the laboratory reference system $\{X,Y,Z\}$ (see Fig. \ref{fig-simgeom}) is as follows:
\begin{itemize}
\item $z_p$ is in the direction of particle motion $\vec{v}=(v_1, v_2, v_3)$, $\vec{v}$ is normalised
\item $x_p$ is parallel to $XZ$ plane
\item $Y$ is parallel to $y_pz_p$ plane.
\end{itemize}
The particle basis definition can be expressed using the formulae
\begin{eqnarray}\label{eq-basis}
  \hat{x}_p & = & (\cos\phi,0,-\sin\phi),\nonumber\\
  \hat{y}_p & = & (-\sin\phi\sin\theta, \cos\theta, -\cos\phi\sin\theta),\\
  \hat{z}_p & = & (v_1,v_2,v_3),\nonumber
\end{eqnarray}
where
\begin{eqnarray*}
\sin\phi & = & \frac{v_1}{\sqrt{v_1^2+v_3^2}},\\
\cos\phi & = & \frac{v_3}{\sqrt{v_1^2+v_3^2}},\\
\sin\theta & = & \frac{v_2}{\sqrt{v_1^2+v_2^2+v_3^2}},\\
\cos\theta & = & \frac{\sqrt{v_1^2+v_3^2}}{\sqrt{v_1^2+v_2^2+v_3^2}}.
\end{eqnarray*}

\subsection{Electrons from \texorpdfstring{$\beta^-$ }{beta- }decay}\label{sec-generation}
The first step of the simulation is the generation of electrons from $\beta^-$ decay with proper
distributions of energy, momentum and polarisation. Although \geant\ provides possibility to generate
decay of neutrons from the beam, it takes so much computational time, that for better efficiency 
the simulation starts from electrons. The neutron beam, however, together with the $\beta$ decay
and neutron physics is implemented in the program and will be helpful in future for identifying
main sources of background.

\begin{figure}[h]\center
\includegraphics[width=0.49\textwidth,height=0.49\textwidth]{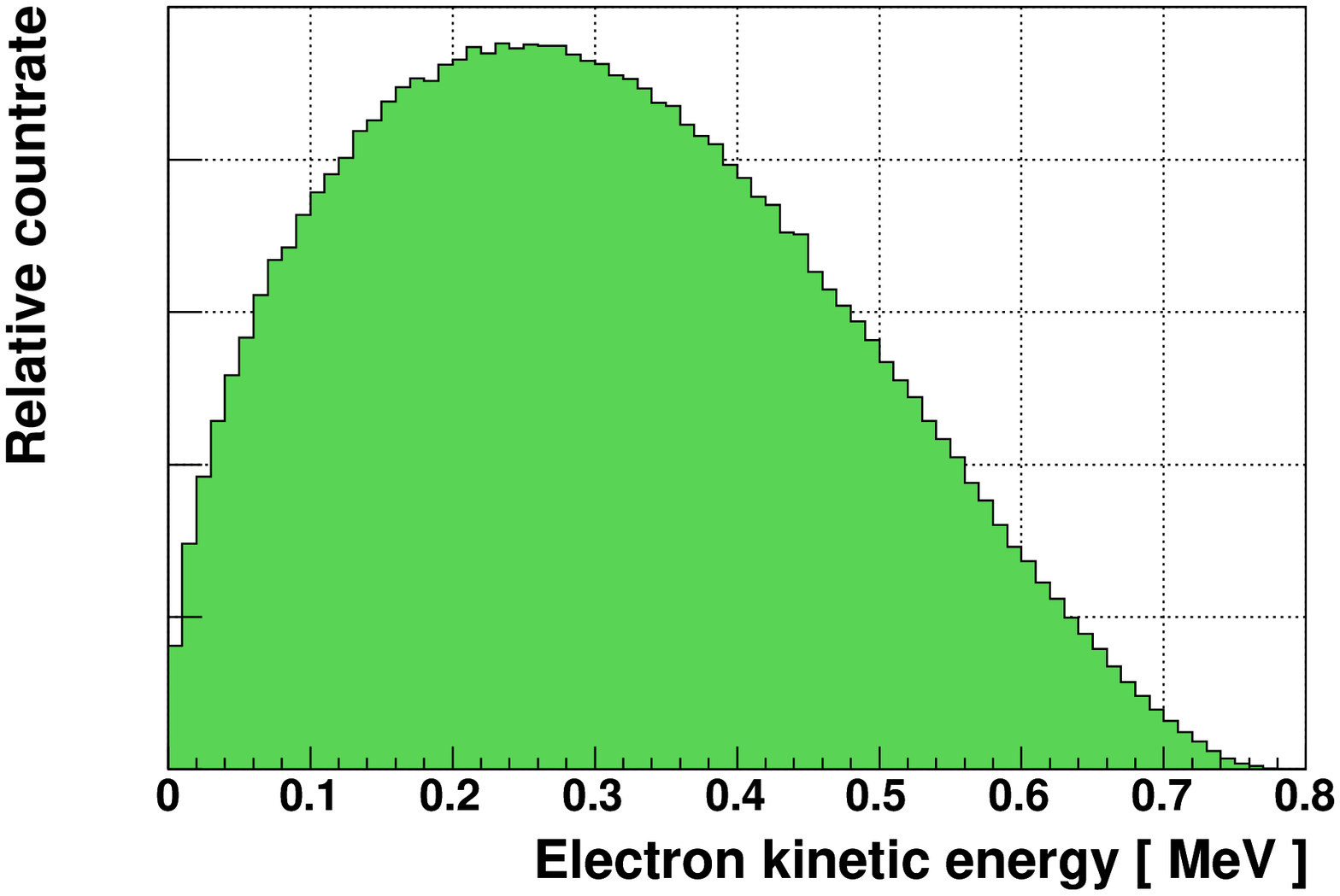}
\includegraphics[width=0.49\textwidth,height=0.49\textwidth]{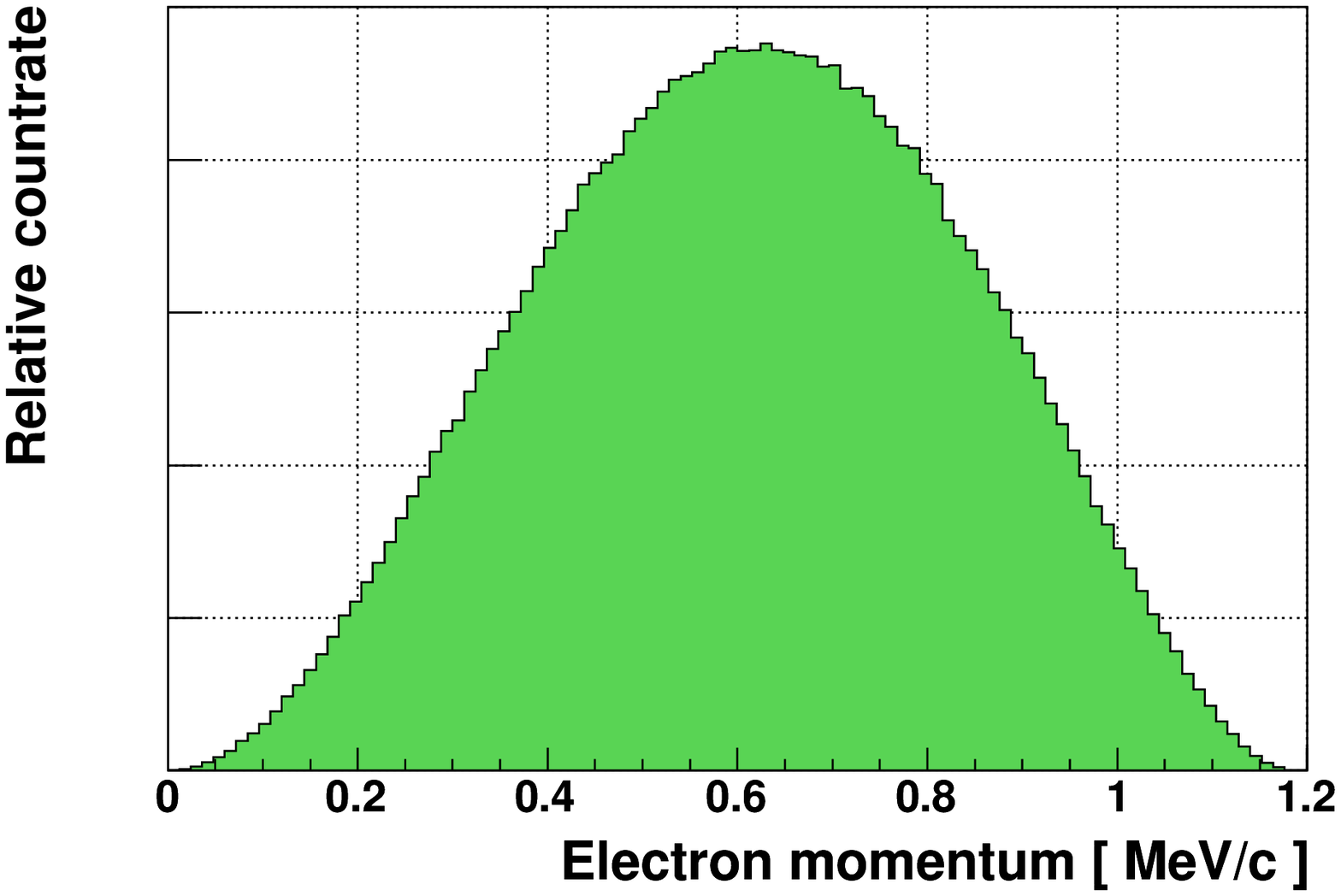}
\caption{Simulated electron energy and momentum spectra (for 1 million events).\label{fig-betaspectra}}
\end{figure}

Electrons are uniformly created inside volume of given position and sizes corresponding to the real 
shape and spatial placement of the neutron beam. Their kinetic energy is sampled using the 
acceptance-rejection method\footnote{Also known as the {\em von Neumann method}\/.}
and an approximated version of Eq. \ref{eq-beta1} (see Fig. \ref{fig-betaspectra})
\begin{equation}
W(E_e) \propto p_eE_e(E_{max}-E_e)^2.
\end{equation}
Finally, 
momentum and polarisation are generated from the probability distribution given by 
\begin{equation}\label{eq-w2}
W\propto\left[1+
  A\frac{\vec{\left<J\right>}\cdot\vec{p}_e}{E_e}+
  G\frac{\vec{\sigma}\cdot\vec{p}_e}{E_e}+
  N\vec{\sigma}\cdot\vec{\left<J\right>} +
  Q\frac{\vec{\sigma}\cdot\vec{p}_e}{E_e+m}\frac{\vec{\left<J\right>}\cdot\vec{p}_e}{E_e} +
  R\frac{\vec{\left<J\right>}\cdot\left(\vec{p}_e\times\vec{\sigma}\right)}{E_e}\right],
\end{equation}
where the polarisation vector $\vec{\sigma}$ is normalised to unity.
It is worth mentioning, that values of the decay parameters together with the mean beam
polarisation $\vec{\left<J\right>}$ can be easily modified by the user (in the file {\tt run.mac}). 
In addition to this, user defined limits on electron energy and emission angle can be applied which 
is necessary for optimisation purposes.

\begin{figure}[!h]
  \center
  \includegraphics[width=0.49\textwidth,height=0.49\textwidth]{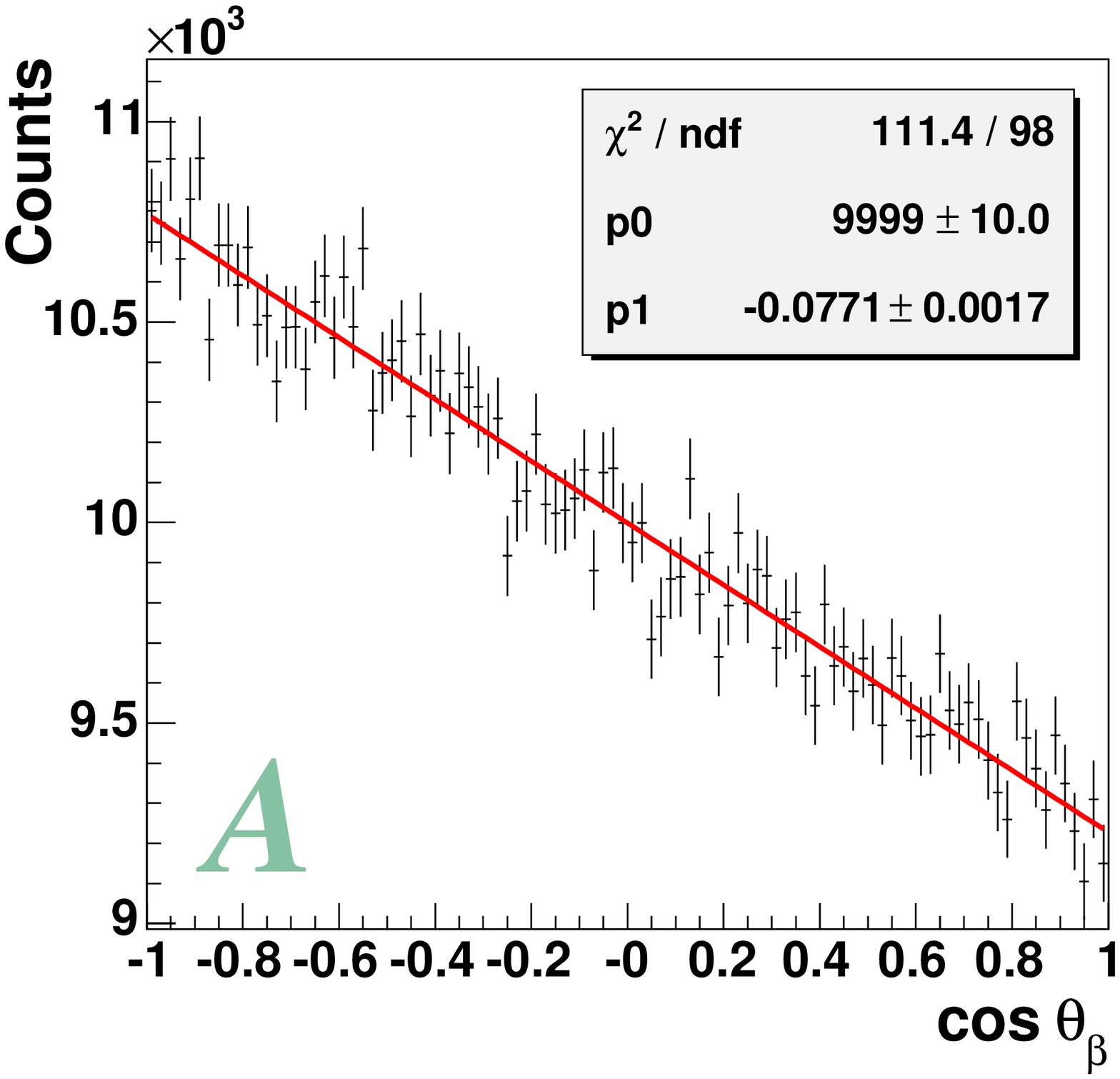}
  \includegraphics[width=0.49\textwidth,height=0.49\textwidth]{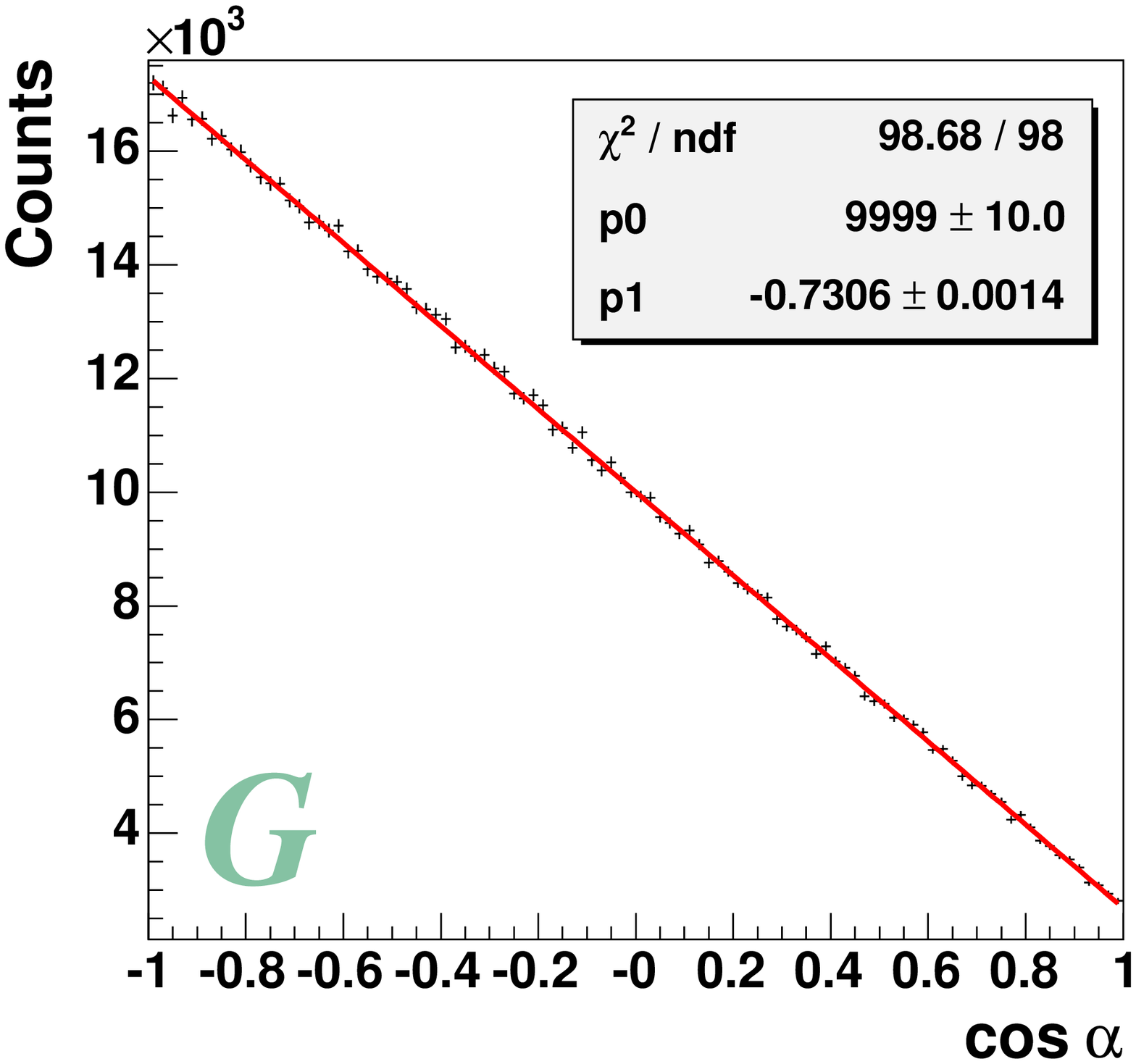}
  \caption{Simulated $A$ and $G$ decay asymmetries with fitted functions \mbox{$y=p_0(1+p_1\cdot x)$}. 
    $\alpha$ is the angle between the electron momentum and polarisation.
    The number of generated electrons equals 1 million.\label{fig-thetabeta}}
\end{figure}

Figure \ref{fig-thetabeta} shows results of a simple test which was performed to check if assumed 
$A$ and $G$ values were reproduced in generated data. Comparing the probability distribution \ref{eq-w2}
with the fitted linear function, one can see that 
\[A_{sim.}=\frac{E_e}{p_e\left<J\right>}\cdot p_1 \mbox{\ \ \ and \ \ \ } G_{sim.}=\frac{E_e}{p_e}\cdot p_1,\] 
where $\left<J\right>=0.8973$ is the mean neutron beam polarisation and $\frac{E_e}{p_e}$ has to be
averaged over the $\beta$ decay energy spectrum (what gives $\left<\frac{E_e}{p_e}\right>=1.368$). 
Finally, one obtains:
\[A_{sim.} = \frac{1.368}{0.8973}\cdot  p_1 = -0.118 \pm 0.003 \mbox{\ \ \ and \ \ \ } G_{sim.} = -0.999 \pm 0.002,\]
which is in perfect agreement with the taken values: $A=-0.1162$ and $G=-1$.
\begin{figure}[!h]
  \center
  \includegraphics[width=0.49\textwidth,height=0.49\textwidth]{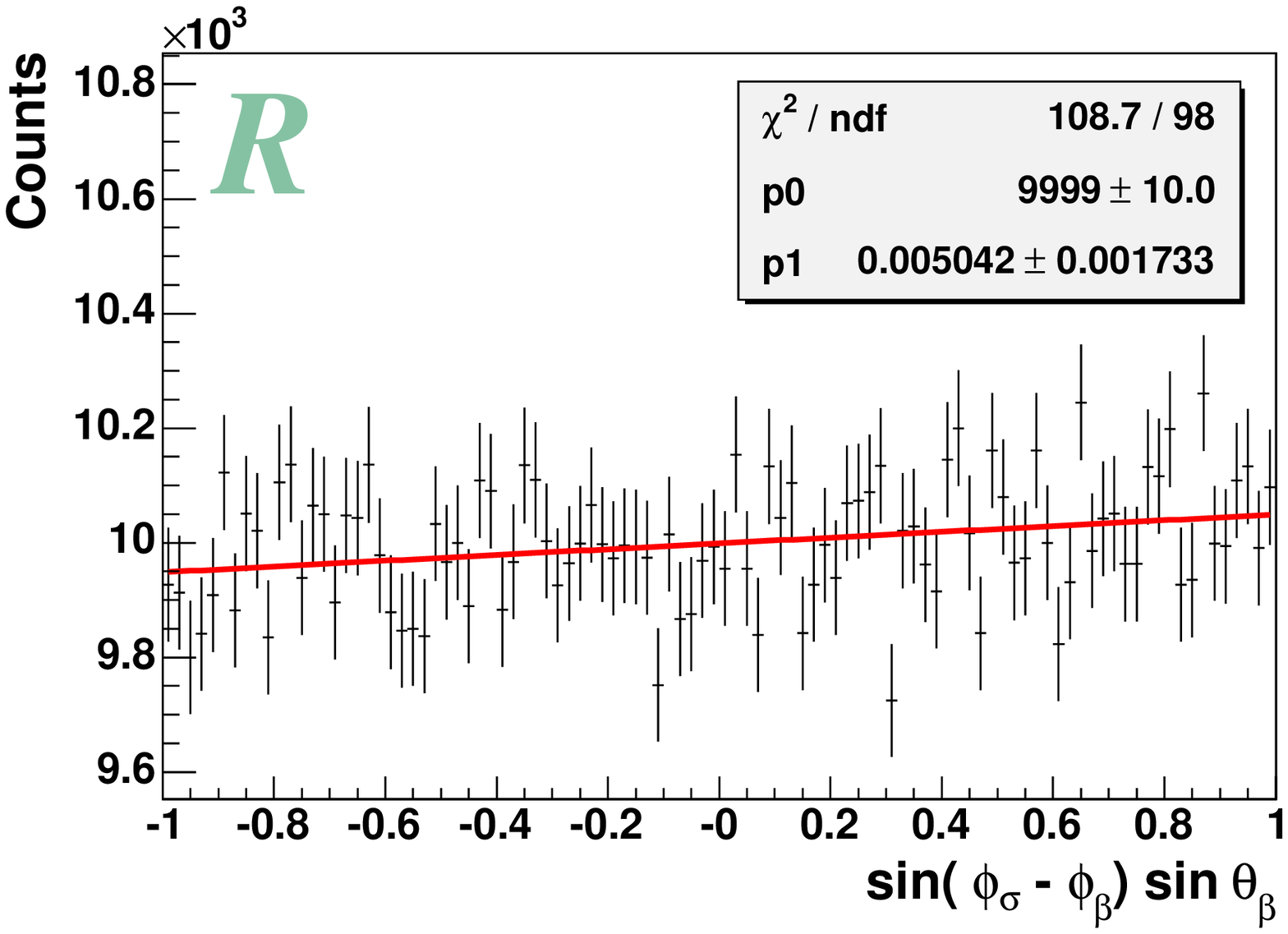}
  \includegraphics[width=0.49\textwidth,height=0.49\textwidth]{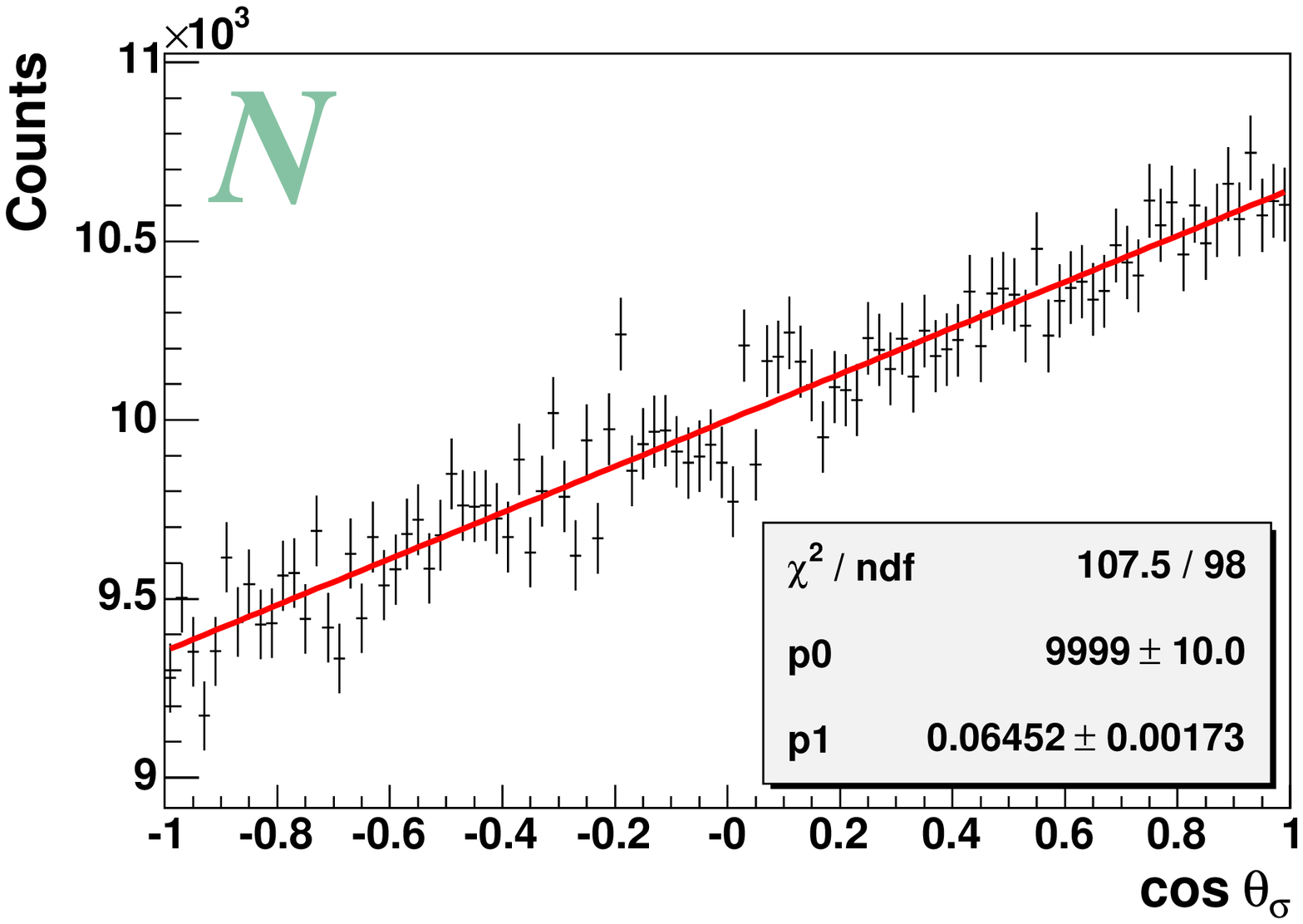}
  \caption{Simulated $R$ and $N$ coefficients. In both cases the fitted function is \mbox{$y=p_0(1+p_1\cdot x)$}. \label{fig-sigma}}
\end{figure}Similar 
tests have been done for $N$ and $R$ asymmetries in the generated spin of the electron.
Again, comparing the formula \ref{eq-w2} with Fig. \ref{fig-sigma} after some calculations, 
one can obtain their values straight from the fit. The results are:
\begin{itemize} 
\item for assumed $R=0.010$
  \[R_{sim.} = \frac{1.368}{\left<J\right>}\cdot  p_1 = 0.008 \pm 0.003,\] 
\item for assumed $N=-A\frac{m_e}{E_e}=0.076$
  \[N_{sim.} = \frac{1}{\left<J\right>}\cdot  p_1 = 0.072 \pm 0.002.\]
\end{itemize}
It is thus confirmed, that $\beta$ decay parameters are reproduced in generated data with
fair accuracy.
\subsection{Depolarisation}
\begin{figure}[!h]
  \center
  \includegraphics[scale=0.6]{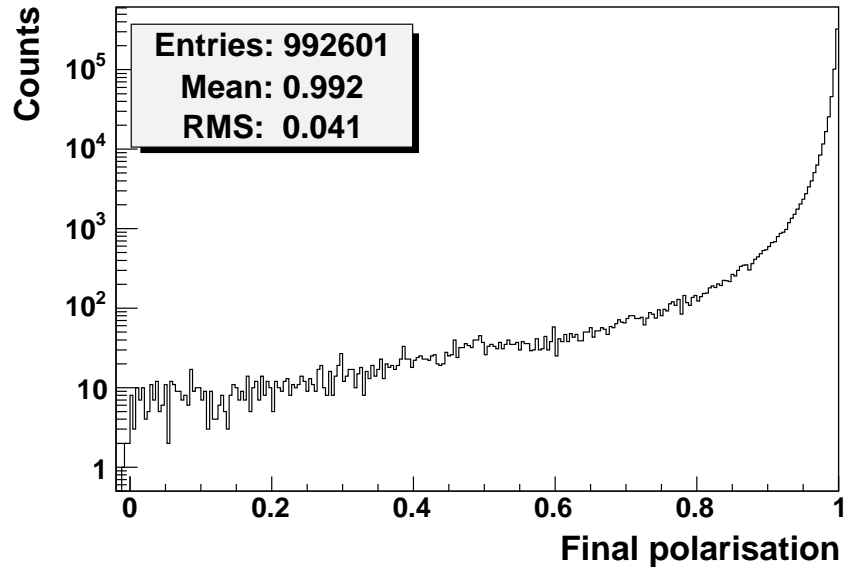}
  \caption{Depolarisation due to the M\o ller scattering\label{fig-moller}}
\end{figure}
\begin{figure}[!h]
  \center
  \includegraphics[scale=0.6]{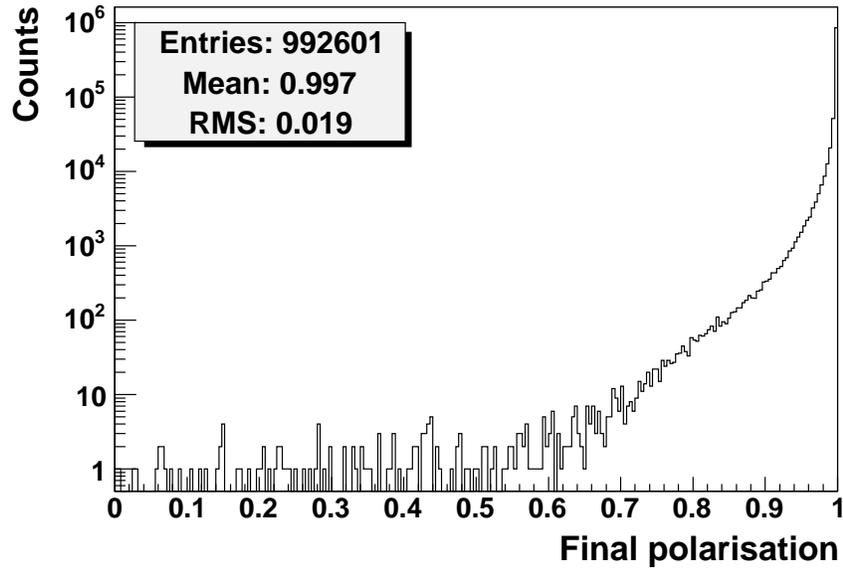}
  \caption{Depolarisation due to the multiple scattering}
\end{figure}
\begin{figure}[!h]
  \center
  \includegraphics[scale=0.6]{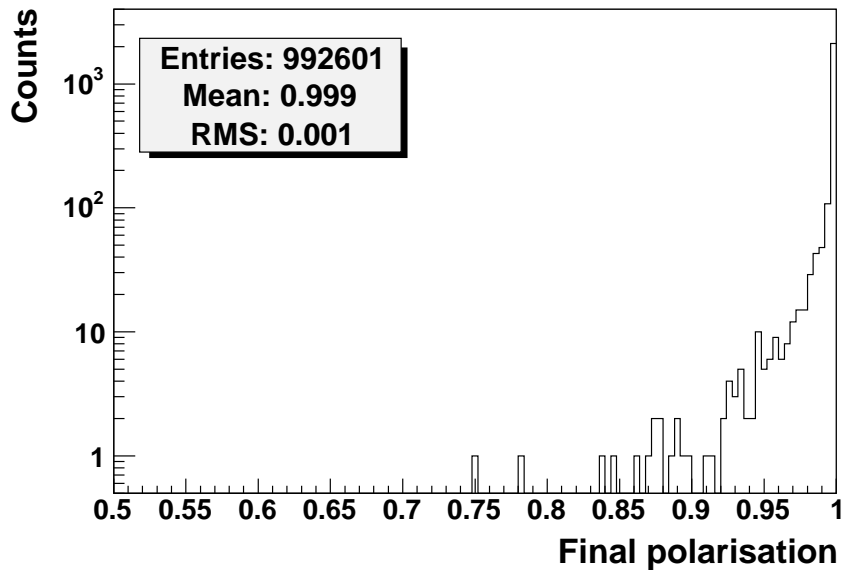}
  \caption{Depolarisation due to bremsstrahlung\label{fig-brem}}
\end{figure}
Generated $\beta$ decay electrons traverse the experimental setup, namely the helium box and its
mylar window, the proportional chamber filled with special gas mixture, another mylar window and the box containing the Pb foil, also 
filled with helium, where their transversal polarisation is to be analysed by Mott scattering. However, before reaching the analysing 
foil, electrons are partially depolarised through numerous interactions. This subsection contains some quantitative estimations of 
the polarisation loss based on the simulation results. In order to respect spin effects, \geant\ libraries have been extended using 
formulae from section \ref{sec-passage}. 
Before the actual results have been acquired, an especial effort had been made to check the reliability of the program and to verify
the outcome with Ref.~\cite[section 4.3]{hoogduin}.

The figures below (\ref{fig-moller} -- \ref{fig-brem}) show the depolarisation effects due to three 
implemented processes, separately (\geant\ gives possibility to ``turn off'' chosen physical 
processes). One million fully transversely polarised electrons have 
been shot from 
the point exactly in the middle of the helium box with momentum vector perpendicular to the Pb 
foil surface and the kinetic energy set to 700~\keV. The polarisation vector in the {\em particle
frame}\/ was $\vec{P}=(1,0,0)$ in order to investigate behaviour of the polarisation
component strictly related to the $R$-correlation. The ``final polarisation'' on the  
pictures is the $\vec{P}_x$ component at the surface of the analysing foil, after the 
transportation through the whole setup.
As it was expected, ionisation is the dominant process and the least change is caused by 
bremsstrahlung.

\subsection{Mott scattering in the analyzing Pb foil}\label{sec-mottexp}
\begin{figure}[!h]
  \center
  \includegraphics[scale=0.82, trim=0 0 100 0]{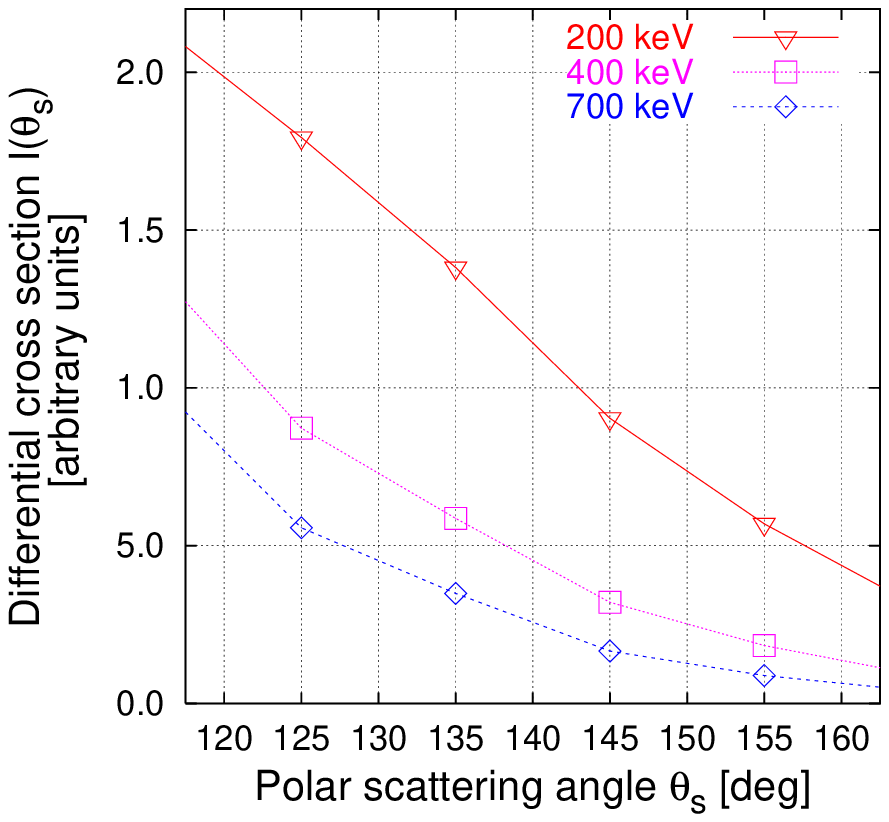}
  \includegraphics[scale=0.82, trim=0 0 100 0]{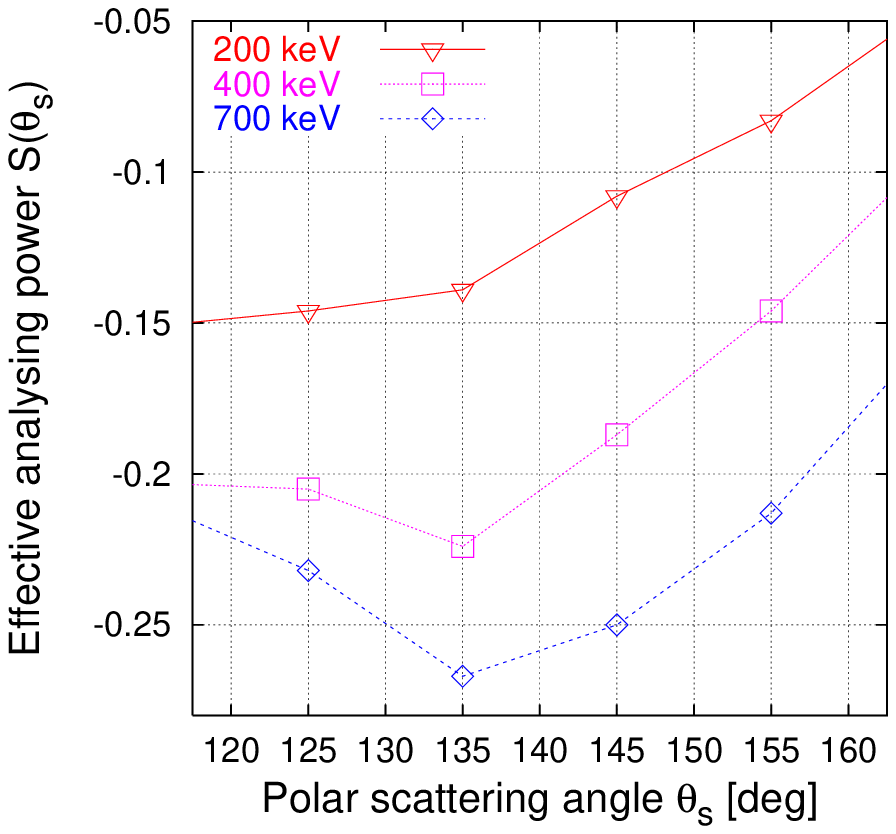}
  \caption{The differential cross section and the effective Sherman function used in the program. Both are results of an independent 
    simulation performed by E.~Stephan~\cite{ela}.\label{fig-seff}}
\end{figure}

After crossing the proportional chamber, a partially depolarised electron can hit the analyzing foil covered with a layer of Pb, 
where it is scattered (as described in section \ref{sec-mott}). The formula \ref{eq-sherman}, which governs this process, can be
rewritten in a form
\begin{equation}\label{eq-sherman2}
\frac{d\sigma}{d\Omega}(\theta,\phi)=
I(\theta)\left[1+S\left(\theta\right)\left(P_y\cos\phi-P_x\sin\phi\right)\right],
\end{equation}
which makes the dependence on the transversal polarisation component more obvious. 
As it was already mentioned, \geant\ does not
contain any electron polarisation dependent processes, therefore to include the Mott scattering some extensions of the code have been 
done. The most important steps of the implemented routine are:
\begin{enumerate}
\item Functions $I$ and $S$ are loaded from the file (see Fig. \ref{fig-seff}). 
\item As soon as an electron reaches the surface of the analyzing foil, the \geant\ tracking engine is stopped.
\item An exact point of the scattering inside the foil is generated from the uniform distribution.
\item The energy loss between the foil border and the scattering point is calculated using the database \cite{eloss} and subtracted from 
  the electron energy.
\item The new particle momentum is generated from the formula \ref{eq-sherman2}. If needed, the values of functions 
  $I$ and $S$ are linearly interpolated.
\item The energy loss between the scattering point and the foil border is calculated and subtracted.
\item The new particle momentum and energy are passed to the \geant\ tracking engine, which is started again.
\end{enumerate}

And here appears an important issue. The formula \ref{eq-sherman2} is valid for the scattering on a single nuclei, what corresponds
to the {\em exact point of the scattering} which is sampled in the third step of the routine. However, the electron in the foil can
be also scattered two or more times, what cannot be neglected, even though for the foil thickness used in the experiment the single
scattering is still more probable (the thicker is the Pb layer, the more probable is double or multiple scattering). The solution
of this problem are, calculated for a given foil thickness, {\em effective} functions $I_{\mbox{\it eff}}(\theta)$ and 
$S_{\mbox{\it eff}}(\theta)$ which 
already contain multiple scattering effects and ought to be used instead of the normal cross section and analysing power.
Both effective functions have been calculated by E.~Stephan \cite{ela} with a dedicated Monte Carlo simulation based on 
Ref.~\cite{hnizdo} and are presented in Fig.~\ref{fig-seff}.

\begin{figure}[!h]
  \center
  \includegraphics[width=0.49\textwidth,height=0.49\textwidth]{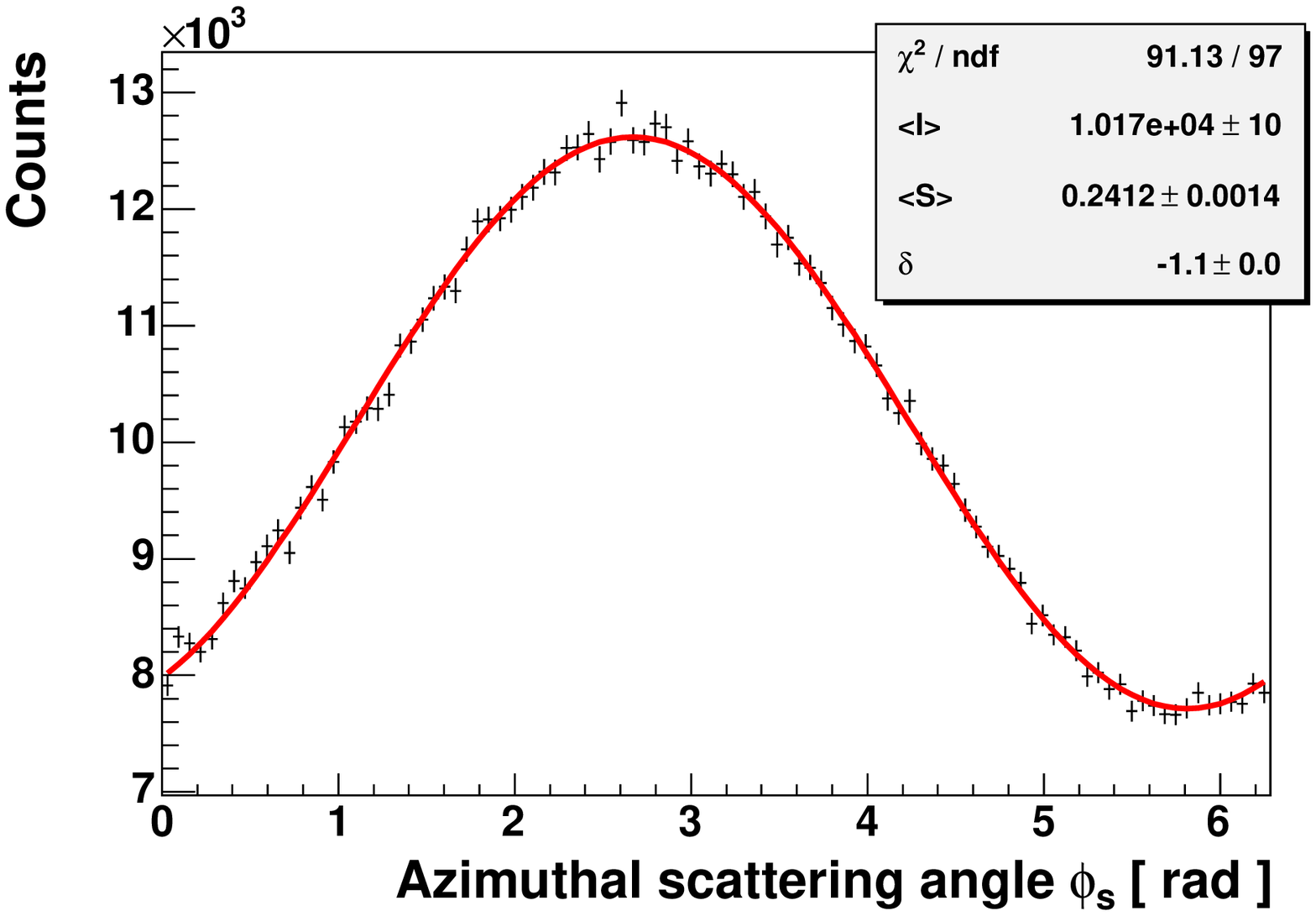}
  \includegraphics[width=0.49\textwidth,height=0.49\textwidth]{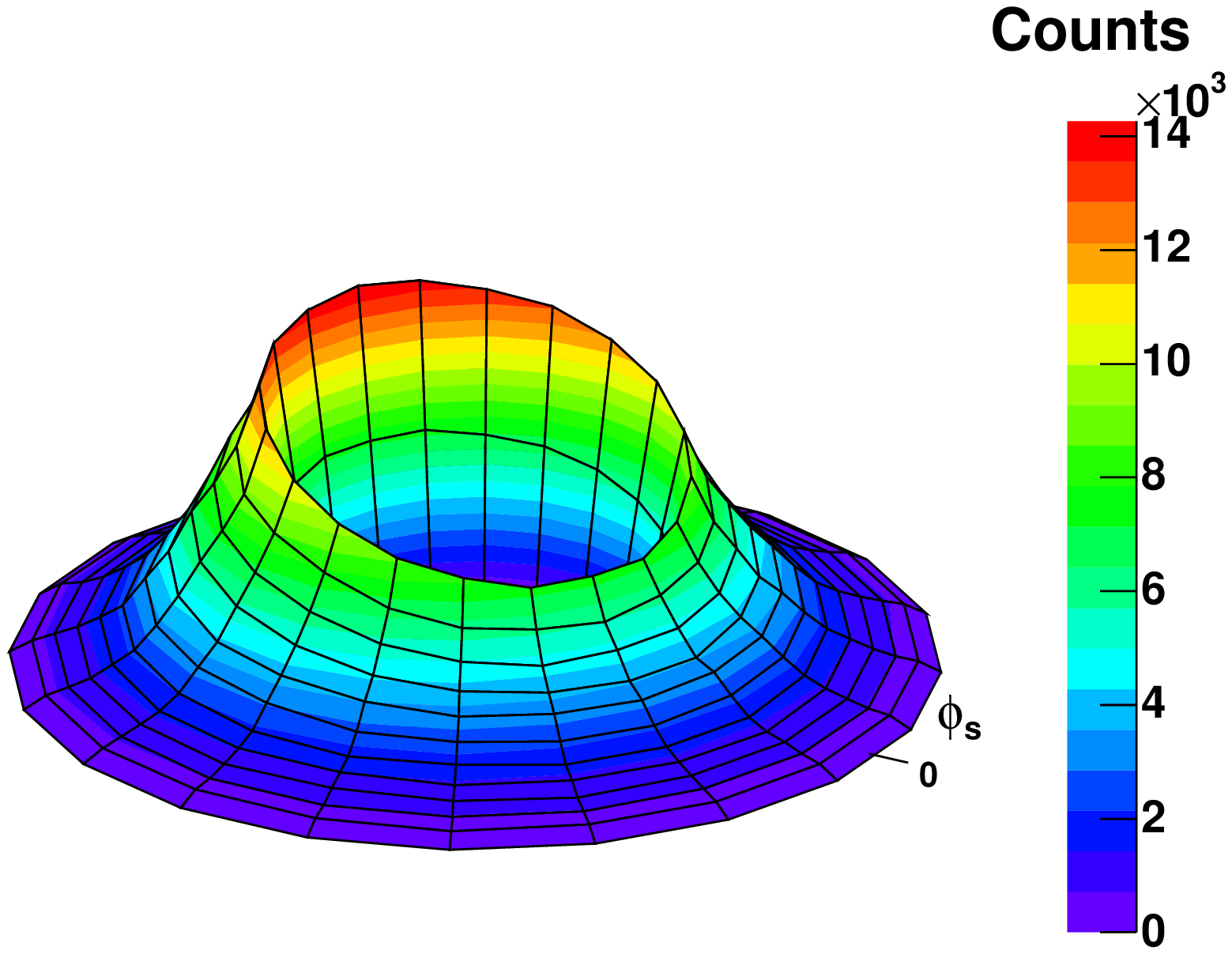}
  \caption{Test of the Mott scattering routine: distributions of the scattering angles. The function fitted to the $\phi_s$ distribution
    is \mbox{$y=\left<I\right>[1+\left<S\right>\sin(\phi_s+\delta)]$}. \label{fig-testmott}}
\end{figure}

In order to check reliability of the Mott scattering routine a simple test have been prepared. Scattering of one million fully transversely 
polarized electrons with the polarisation vector
\[ \vec{P} \equiv (P_x,P_y,P_z) = \left(\frac{1}{\sqrt{5}}, \frac{2}{\sqrt{5}}, 0\right) = (0.447, 0.894, 0)\] 
and kinetic energy 700~\keV\ have been simulated, resulting in angular distributions shown on Fig.~\ref{fig-testmott}. The scattering angle
$\theta_s$ here, and for all other simulations, was restricted to the region 120 -- 160 degrees, because for lower angles reflected electrons 
do not have a chance to hit the detector and for larger angles the cross section goes down rapidly. The cross section \ref{eq-sherman2} 
averaged over the $\theta_s$ angle can be written as
\begin{equation}\label{eq-sherman3}
\frac{d\sigma}{d\Omega}(\phi_s) =
\left<I\right>\left[1+\left<S\right>\sin\left(\phi_s+\delta\right)\right],
\end{equation}
therefore full information about the transversal polarisation component is hidden in the phase shift $\delta$ and can be obtained 
from the fit (see Fig.~\ref{fig-testmott}). The results
\begin{eqnarray*}
P_x = -\cos\delta = 0.454 \pm 0.006 & & P_y =  \sin\delta = 0.891 \pm 0.003,
\end{eqnarray*}
are in a very good agreement with assumed values, what provides a satisfactory crosscheck.

One should be aware of several systematic errors that might be introduced here and could lower the accuracy of the simulation, especially
when it will be used for prediction of the real measurement results:
\begin{itemize}
\item the effective functions $I_{\mbox{\it eff}}$ and $S_{\mbox{\it eff}}$ are model dependent and can contain some
  systematics caused by the generation method (\cite{hnizdo}),
\item because of the unavailability of proper data for $_{82}$Pb, the analyzing power $S$ for $_{80}$Hg has been used as the input for 
  the $S_{\mbox{\it eff}}$ calculation,
\item small additional deviations are caused by the linear interpolation of $I_{\mbox{\it eff}}$ and $S_{\mbox{\it eff}}$
  values between the points of given energy and scattering angle,
\item so far the electron depolarisation inside the Pb-foil, directly before the scattering has not been implemented.
\end{itemize}
\cleardoublepage
\chapter{Systematic effects}
After numerous careful tests of the program, described in the previous chapter, it finally could have been employed in  
realistic simulation of the whole experiment and its conditions. The results presented below required two days of 
calculations on a PC machine with a fast CPU. It was enough to reach the expected statistics of the planned experiment, 
namely, over one million events (V-Tracks). Due to the significant differences in cross sections and analysing powers, the 
data analysis was performed separately for three electron energy intervals: 
\begin{itemize}
\item 200 \keV\ -- 400 \keV,
\item 400 \keV\ -- 600 \keV,
\item 600 \keV\ -- 800 \keV.
\end{itemize}
Electrons with energies below 200~\keV\ were not generated, since the threshold of our apparatus is expected to be around 
this value. For optimisation reasons, after dedicated tests, some cuts have been also applied on electron generation angles:
\begin{itemize}
\item the polar angle: $\theta_\beta\in(15\degree,165\degree)$,
\item the azimuthal angle: $\phi_\beta\in(20\degree,160\degree)$.
\end{itemize}
Moreover, the Mott scattering angle $\theta_s$ was restricted to a subset $(120\degree,160\degree)$. The motivation is clear, 
on the one hand side, there is no chance to detect in the opposite chamber an electron scattered with $\theta_s<120\degree$, on the
other hand for the scattering angles over $160\degree$ electron tracks before and after scattering are not distinguished by 
the wire chamber.

\section{Data analysis}
First, it is necessary to describe the general concept of the data analysis. For each Mott scattering event (V-Track
\footnote{An electron scattered from the foil which hits the scintillator on the opposite side of the beam}) the simulation saves to 
a file the data specified below:
\begin{itemize}
\item initial momentum direction,
\item momentum at the Pb foil surface, before the scattering,
\item momentum after scattering,
\item initial polarisation,
\item polarisation at the Pb foil surface, before the scattering,
\item the initial electron position,
\item the scattering point position,
\item final position at the scintillator border,
\item initial kinetic energy,
\item final kinetic energy,
\item random deviation from the final kinetic energy (to simulate the effect of the finite energy resolution),
\item artificially generated response of electrodes from the wire chambers.
\end{itemize}
The raw data files, containing values in the ASCII format, are later converted with the extended version of the {\sc NPRun}\cite{nprun} 
analysis program to binary {\tt *.root} files. During this process a few important parts of the actual analysis are done, including the 
reconstruction of V-Tracks from the artificial wire chamber data. Afterwards one can work directly on the binary files, using the 
{\sc ROOT} \cite{root} environment and macros, created especially for that purpose. At this point, the analysis is done separately for 
momenta and positions reconstructed from artificial tracks and separately for the ``real'' values, taken straight from the 
simulation.

To obtain values of $R$ and $N$ coefficients from the simulated data, one has to know the intensity distribution of the azimuthal 
scattering angle
$\phi_s$, which can be used to calculate mean values of the transversal polarisation components, just as it was shown in section 
\ref{sec-mottexp}. Both scattering angles can be calculated from momentum directions before and after scattering, transformed
to the incoming particle frame (see section \ref{sec-spintrans}). 
The Fig.~\ref{fig-reconst} shows the distribution of scattering angles for over 200000 V-Tracks, generated from the beam polarised 
in the ``up'' direction.
Its shape is mostly a consequence of acceptance of the the experimental setup and beam geometry and in this form cannot be used for 
determination of the electron spin direction. To achieve a distribution independent on the geometry, just like in the real experiment, 
one has to flip the 
beam polarisation and produce the same amount of data. In this case, the influence of the geometry is exactly the same, while the
effects caused by the polarisation contribute to the $\phi_s$ distribution with the
opposite sign (from equations \ref{eq-w2} and \ref{eq-sherman2}). After subtracting both distributions and normalising the result
to the total number of events, one gets the final, geometry independent intensity distribution of Mott scattering angles for V-Tracks. 
The mean values of polarisation components can be now extracted from the fit.

\section{Reconstruction of V-Tracks}
Probably the most challenging problem in the analysis of data from the real experiment is the reconstruction of electron tracks
based on signals from the MWPC chambers. To avoid depolarisation and multiple scattering effects, the chambers are relatively thin
(around 10~cm) and contain only 5 planes of electrodes. Each plane consists of one layer of anodes 
(horizontal wires) and two layers of cathodes (vertical wires). 

The reconstruction algorithm, implemented in the {\sc NPRun} program is in its advanced
development phase and still requires some tests, while the simulation provides the only possibility to compare ``real'' tracks with
the reconstructed ones and to check the reconstruction efficiency. Therefore, some qualitative comparisons have been done.
\begin{figure}
  \center
  \includegraphics[width=0.49\textwidth]{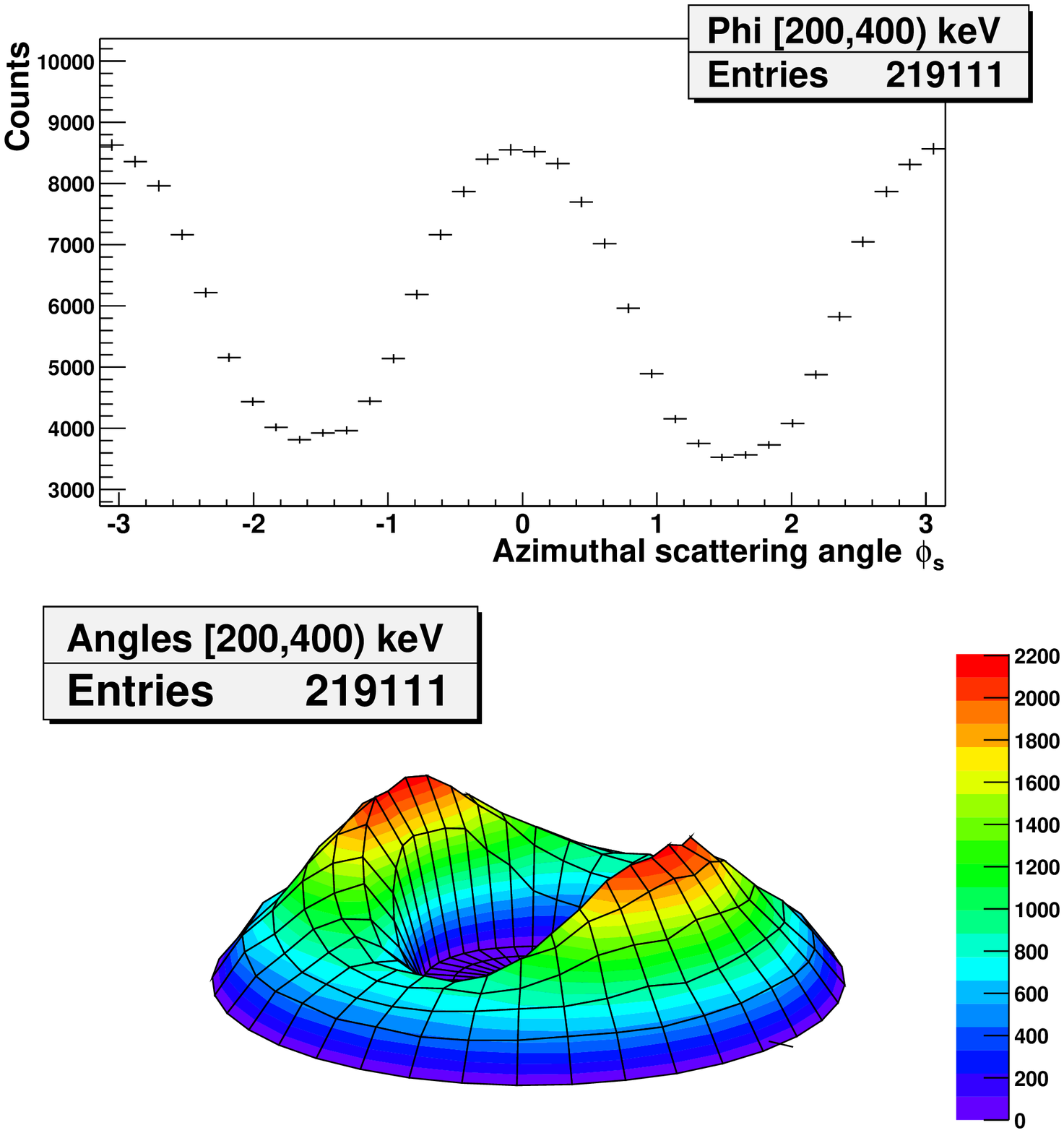}
  \includegraphics[width=0.49\textwidth]{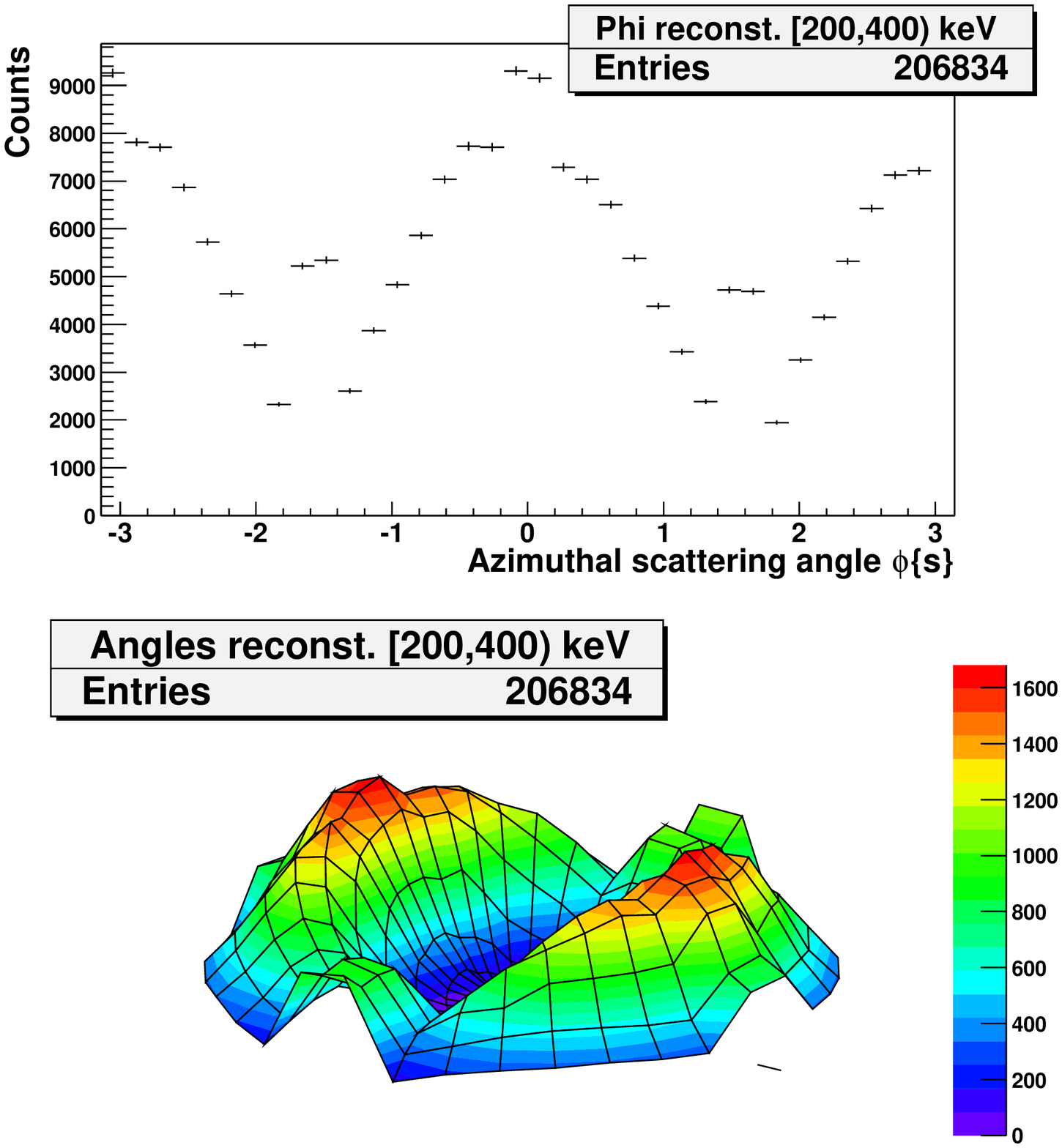}
  \caption{The comparison of the distribution of real scattering angles $\phi_s$ (pictures on the left) with the reconstructed ones 
    (on the right).\label{fig-reconst}}
\end{figure}
As we can see in Fig.~\ref{fig-reconst}, the reconstruction works fine, with very high efficiency, however, 
for the angles $\phi_s=0,\pm\frac{\pi}{2},\pm\pi$ there is a large difference. The missing reconstruction efficiency in that specific
geometry is due to MWPC wire layout feature.
The reconstruction of 
V-Tracks has to be done separately for two projections (for anodes and cathodes). The problem arises for V-Tracks which lay either in 
the cathodes or in the anodes plane. Since they can be fully reconstructed in only one projection, ambiguities can appear. The 
scattering angle values $\phi_s=0,\pm\frac{\pi}{2},\pm\pi$ correspond exactly to these events. The version of the algorithm tested in
this case reconstructs as much of these type of V-Tracks as possible which results in many misidentified tracks.
\begin{figure}
  \center
  \includegraphics[width=0.49\textwidth, height=0.49\textwidth]{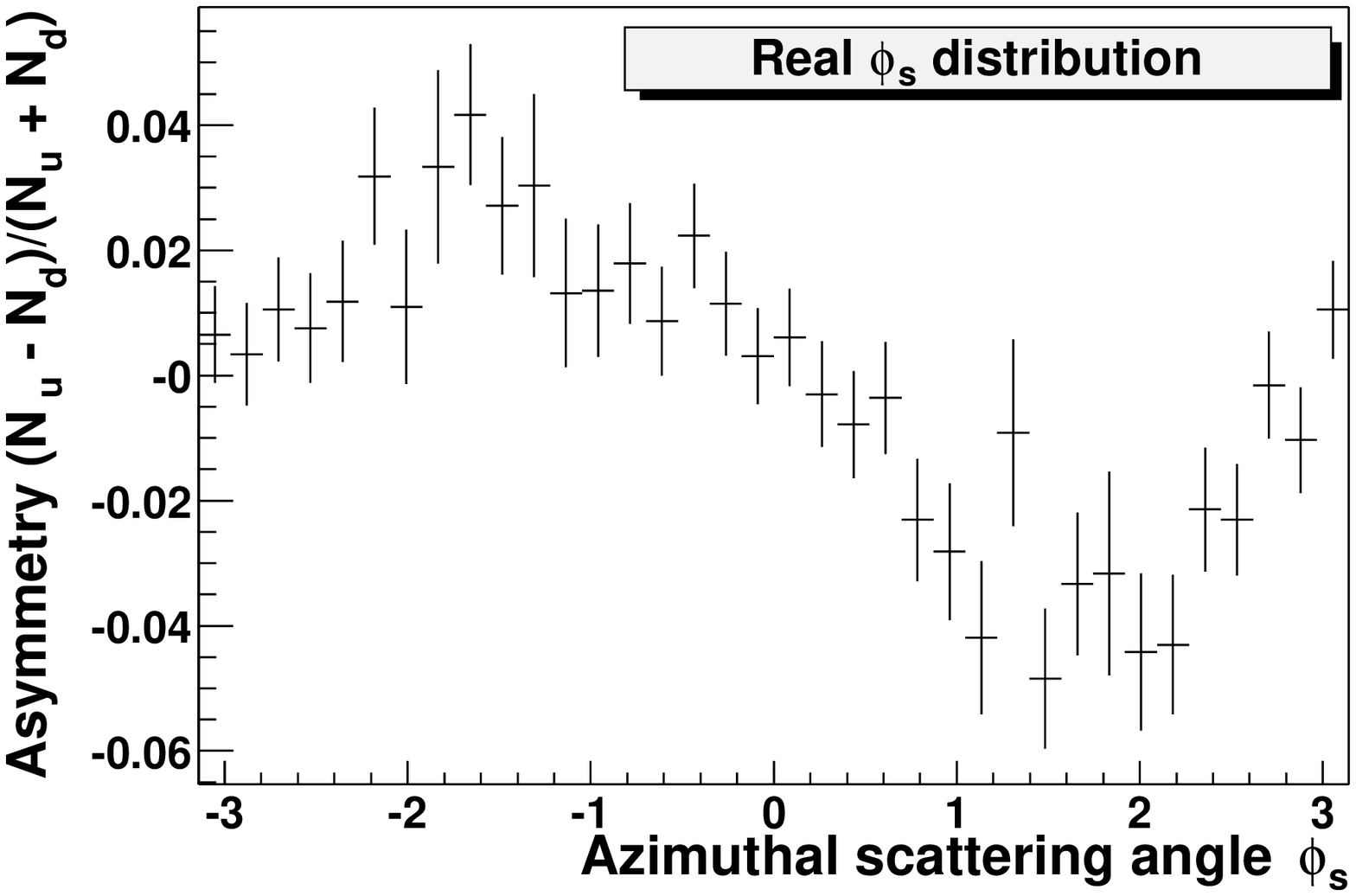}
  \includegraphics[width=0.49\textwidth, height=0.49\textwidth]{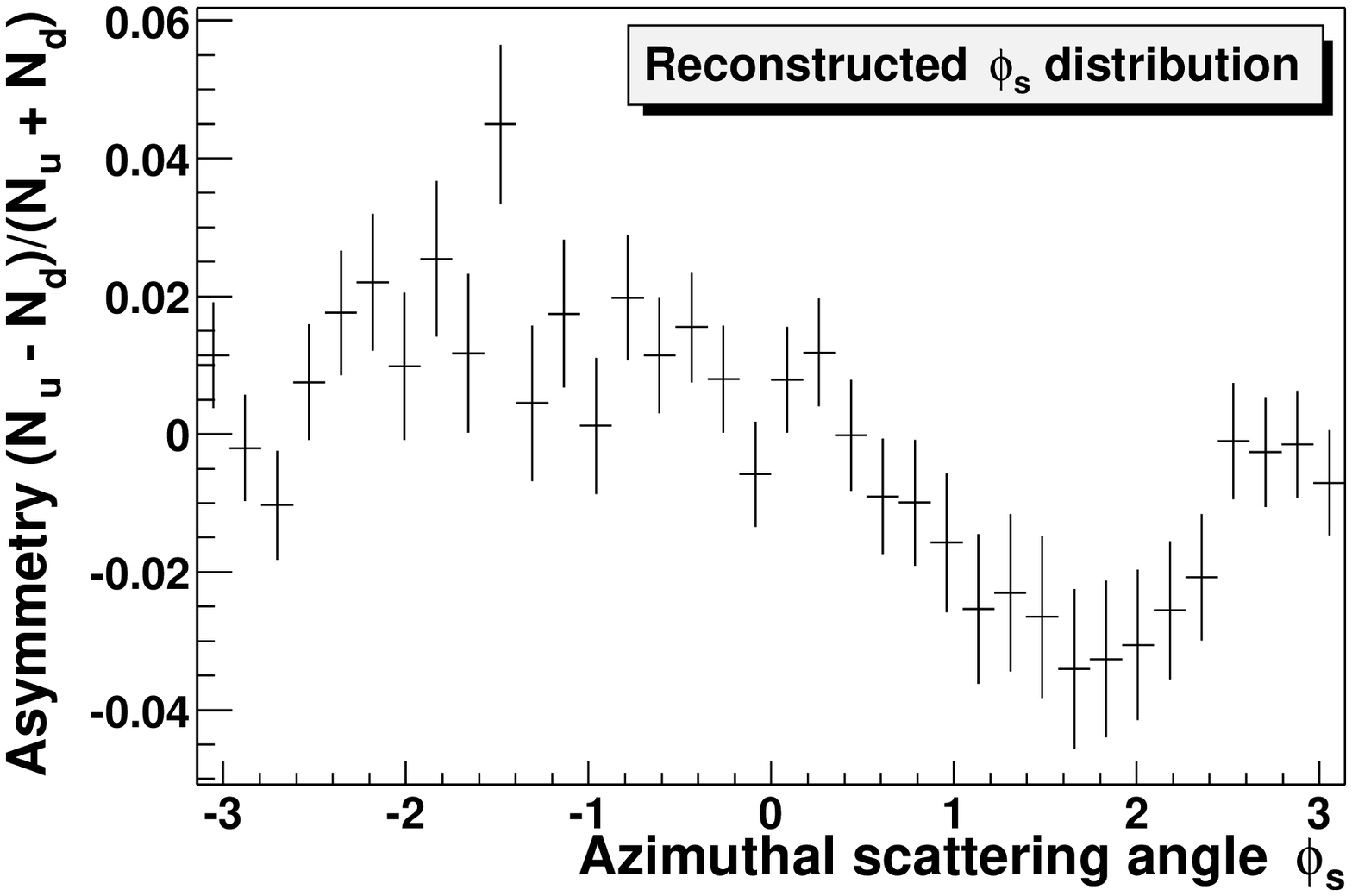}
  \caption{The comparison of the real scattering angles (picture on the left) with the reconstructed ones 
    (on the right) after the subtraction of data corresponding to opposite beam polarisations.\label{fig-reconst2}}
\end{figure}
At the final stage of the analysis the effect cancels out (on the cost of lower statistics) and the ultimate result is 
satisfactory (Fig.~\ref{fig-reconst2}). It 
should be added here, that in reality the efficiency of the reconstruction algorithm is much lower, due to the noisy signal from
the chambers caused by various sources of background. The algorithm for the artificial chamber signal generation
includes the optional noise generation, but in this case it has not been used.

\section{False R-correlation}
\begin{figure}
  \center
  \includegraphics[width=0.7\textwidth]{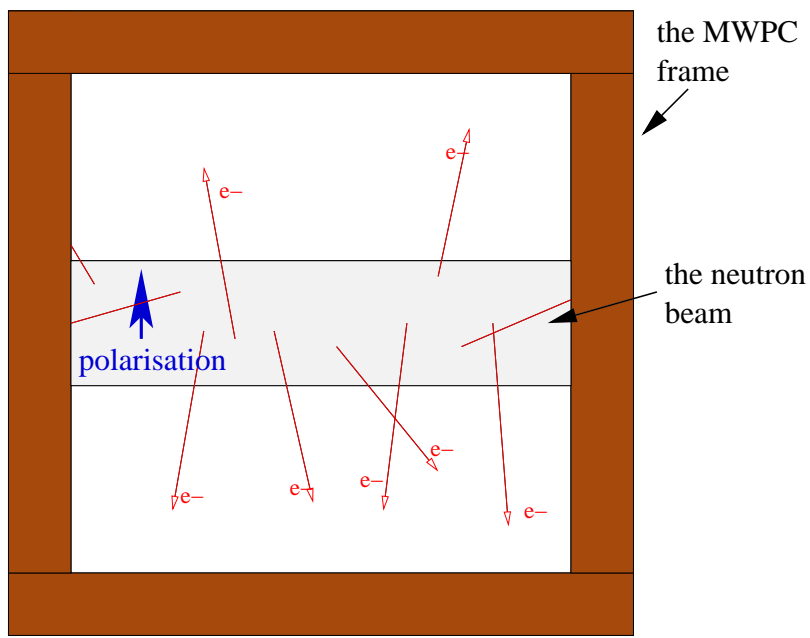}
  \caption{The neutron beam as seen from the position at the analysing Pb foil.\label{fig-Rbeam}}
\end{figure}

\begin{figure}
  \center
  \includegraphics[width=0.49\textwidth]{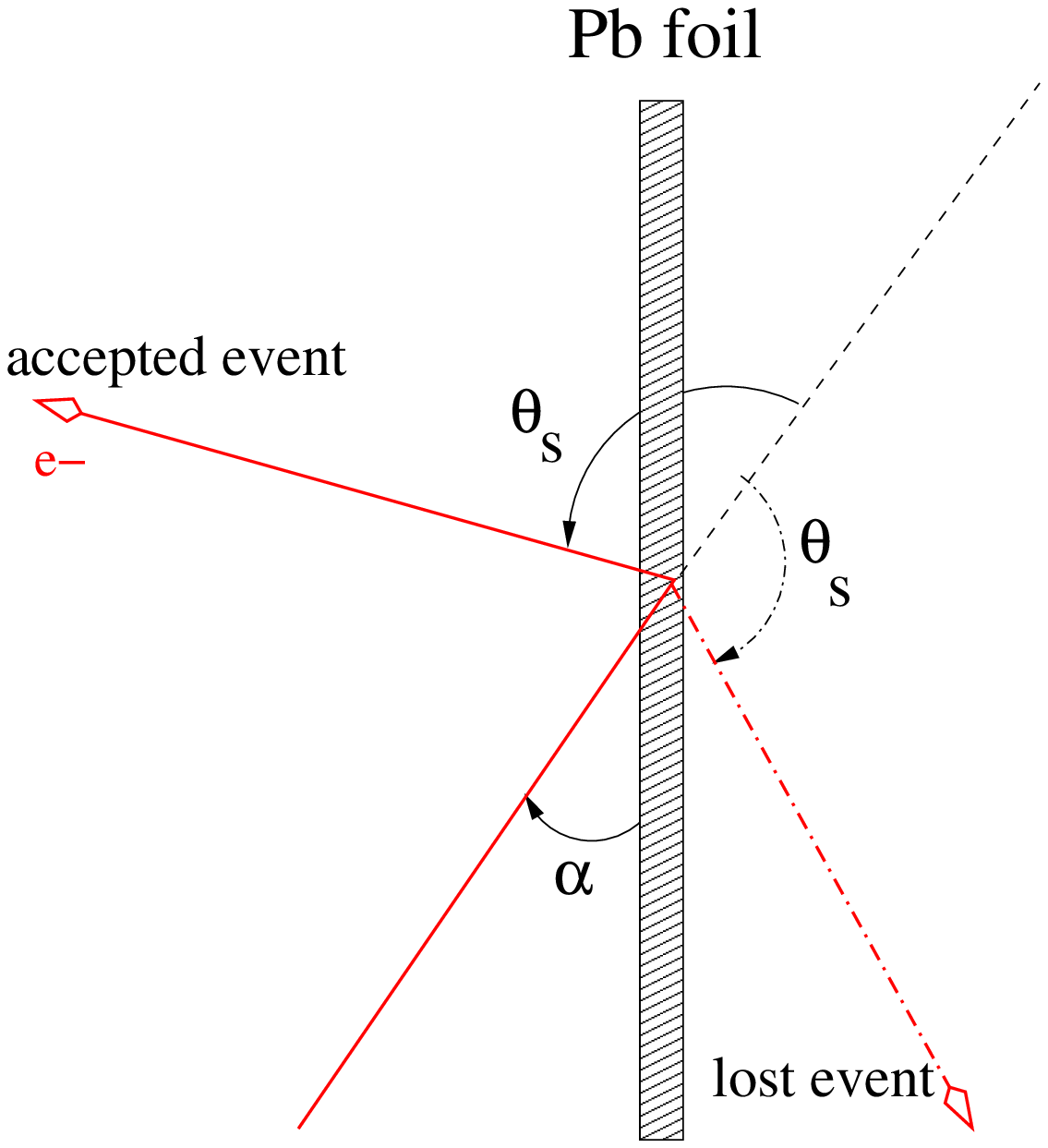}
  \includegraphics[width=0.49\textwidth,height=0.49\textwidth]{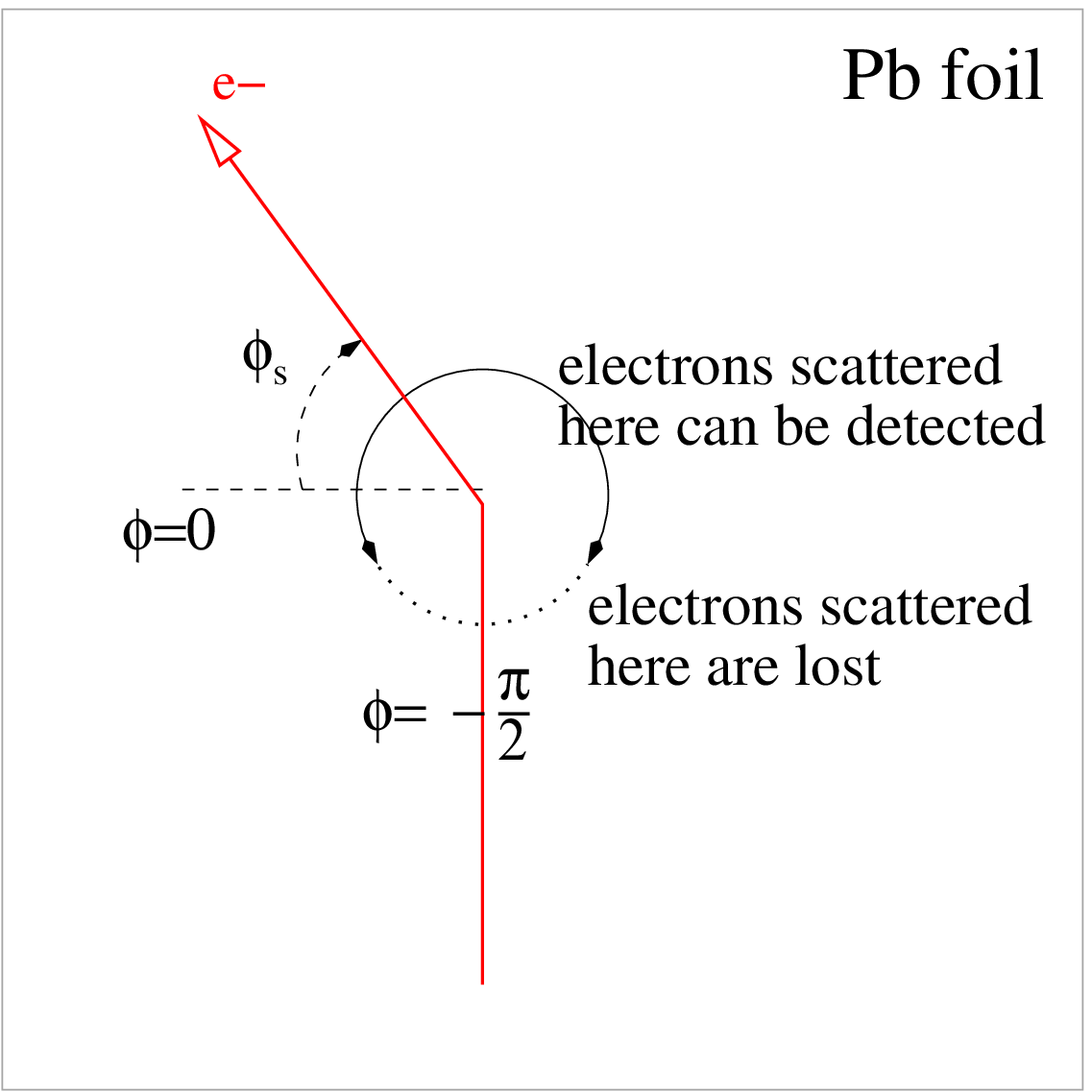}
  \caption{The principle of the systematic effect. For electrons emitted in the direction ``up'' 
    the scattering angle $\phi_s=-\frac{\pi}{2}$ is less probable than the angle $\frac{\pi}{2}.$\label{fig-Rprinciple}}
\end{figure}

\begin{figure}
  \center
  \includegraphics[width=\textwidth]{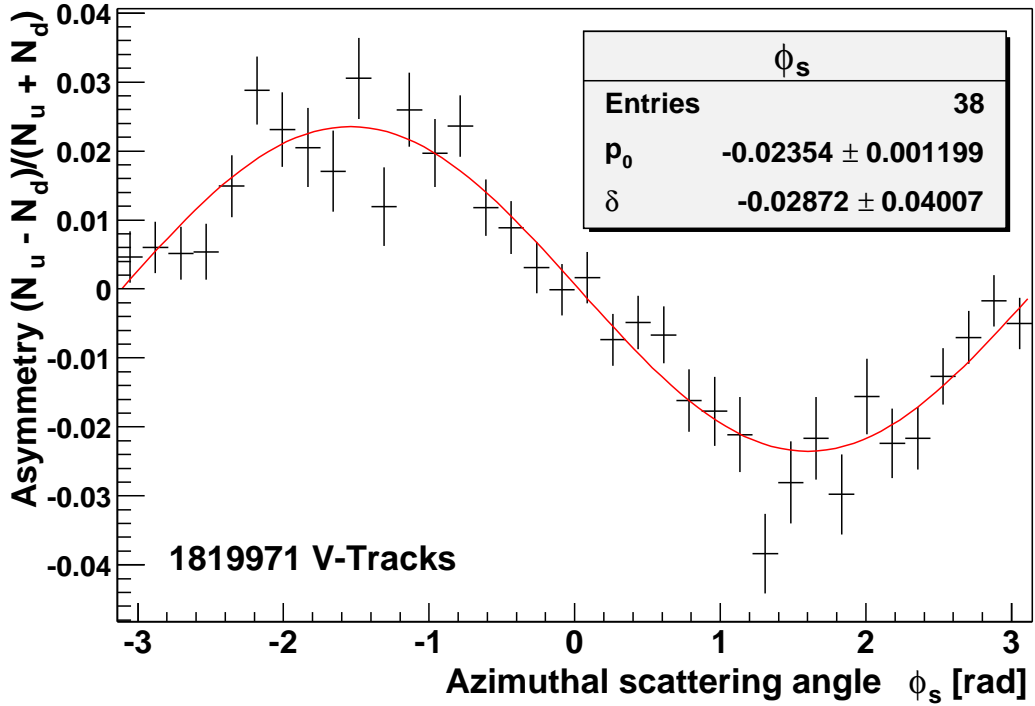}
  \caption{The false $R$ effect for the initial electron energy $E_k\in(200,782) \keV$. $N_u$ and 
    $N_d$ denote the count rate for the ``up'' and ``down'' beam polarisation, respectively. The function
    $y=p_0\cdot\sin(\phi_s + \delta)$ has been fitted to data points. \label{fig-Rsum}}
\end{figure}

Although, the false contribution to the $R$ coefficient caused by the neutron decay asymmetry was one of the expected systematic effects 
(a similar effect appeared in the $R$ measurement for $^8$Li \cite{Li}), its nature have been understood and the magnitude have been 
estimated very recently, by means of this simulation. 
Figures \ref{fig-Rbeam} and \ref{fig-Rprinciple} explain the source of the effect. Let's consider the situation when the beam is
polarised in the ``up'' direction. For simplicity, the dependency on the electron polarisation is not taken into account. Because 
of the neutron decay asymmetry (nonzero $A$ coefficient), more electrons are emitted
``down'', in the direction opposite to the beam polarisation.
In the case of the electrons emitted ``up'', the scattering angle value $\phi_s=-\frac{\pi}{2}$ is less probable than $\frac{\pi}{2}$,
since only the {\em backscattered} particles are accepted by the simulation. Knowing that in the program 
$\theta_s\in(120\degree,160\degree)$, one can see that the effect starts to appear for the angle between the incoming electron
momentum and the scattering foil plane $\alpha<60\degree$. In contrary to this, for the much larger number of particles 
emitted in the direction ``down'', the effect is opposite and $\phi_s=-\frac{\pi}{2}$ is more probable than 
$\frac{\pi}{2}$. 
Of course, if the beam polarisation is flipped, the effect is reversed, therefore it cannot be canceled by subtraction
of $\phi_s$ distributions measured for both beam polarisations. It results in the final sine-like angular distribution,
with one maximum at $\phi_s=-\frac{\pi}{2}$ and one minimum at $\phi_s=\frac{\pi}{2}$ (see Fig. \ref{fig-Rsum}). As it was 
specified before, it is exactly 
the shape that one would expect for the nonzero $R$ coefficient. Therefore, it is necessary to examine this effect and 
estimate the induced false contribution to the measured $R$ coefficient.

\begin{figure}
  \center
  \includegraphics[width=\textwidth, height=0.8\textheight]{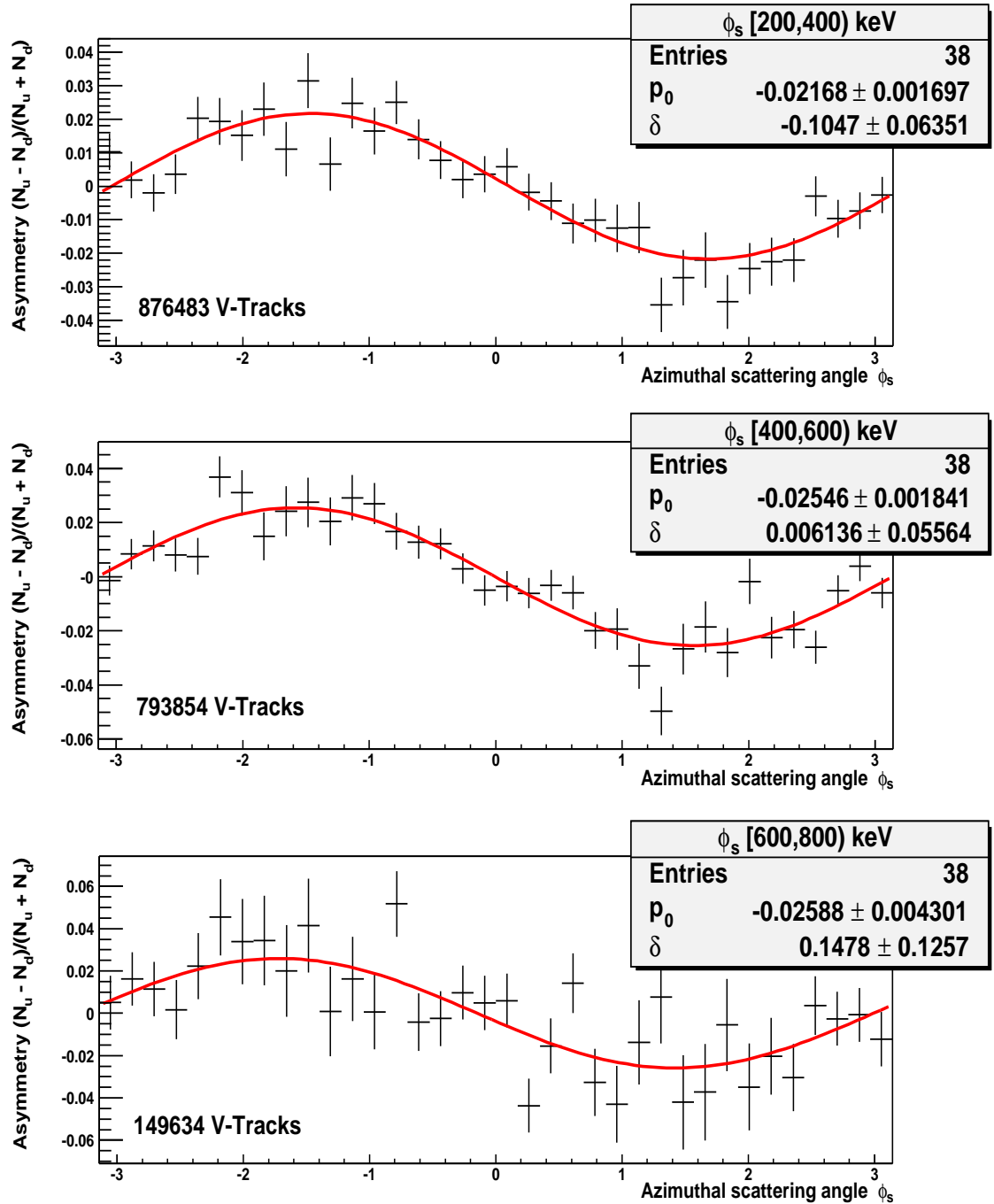}
  \caption{The false $R$ effect for three electron energy intervals. The function
    $y=p_0\cdot\sin(\phi_s + \delta)$ has been fitted to data points. \label{fig-Rclasses}}
\end{figure}

A dedicated simulation have been performed for that purpose. In order to separate the effects induced by the $\beta$ decay 
asymmetry, all the coefficients except $A$ have been set to zero. The program has produced almost 2 million V-Tracks, half of 
them with the beam polarisation ``up'' and half with the opposite. The total result averaged over the electron kinetic energy
$E_k>200~\keV$ and over all generation angles is presented in the Fig. \ref{fig-Rsum}. 
However, even more interesting are similar pictures for three distinct electron energy intervals (Fig.~\ref{fig-Rclasses}).
The amplitudes $p_0$ taken from the fit can be now used to calculate the false contribution to the average value 
of the transversal electron polarisation component. The intensity distributions are given by:
\begin{eqnarray*}
N_u(\phi_s) &=& \left<I\right>[1+\left<S\right>\sin(\phi_s+\delta)]=\left<I\right>[1+\left<S\right>(P_y\cos\phi_s-P_x\sin\phi_s)],\\
N_d(\phi_s) &=& \left<I\right>[1+\left<S\right>(-P_y\cos\phi_s+P_x\sin\phi_s)],\\
\end{eqnarray*}
thus
\[\frac{N_u-N_d}{N_u+N_d} = \frac{2\left<I\right>\left<S\right>(P_y\cos\phi_s-P_x\sin\phi_s)}{2\left<I\right>}=
\left<S\right>\sin(\phi_s+\delta).\]
The average Sherman function values for specified energy ranges are:\\
\begin{center}\begin{tabular}{c||c|c|c}
$E_k [ \keV ]$   & 200 -- 400 & 400 -- 600 & 600 -- 800 \\ \hline
$\left<S\right>$ &   -0.15    &   -0.20    &    -0.23   \\    
\end{tabular}\end{center}
and both transversal polarisation components:
\begin{eqnarray*}
P_x = -\frac{\cos\delta}{\left<S\right>}\cdot p_0, & & P_y =  \frac{\sin\delta}{\left<S\right>}\cdot p_0.
\end{eqnarray*}
The results are:
\begin{center}\begin{tabular}{c||c|c|c}
    $E_k\ [\keV]$   & 200 -- 400          & 400 -- 600        & 600 -- 800         \\ \hline
    $P_x$            & $0.14 \pm 0.01    $ & $0.13 \pm 0.01   $&$  0.11 \pm 0.02$   \\    
    $P_y$            & $0.00 \pm 0.01    $ & $0.00 \pm 0.01   $&$  0.00 \pm 0.02$ \\
\end{tabular}\end{center}
Note, that the systematic error of the average $S$ has not been taken into account in the error estimation.
An obvious conclusion from the above simple analysis is that the magnitude of the investigated systematic effect
does not depend on the electron energy.

It should, however, depend on the electron emission angle as it was explained before. And that was the motivation of the last test 
that has been done. The false $R$ effect has been extracted from the data separately for different values of the polar angle between 
the normal to the scattering foil (the $x$ axis in the main reference system) and the {\em initial} electron momentum. In this case, 
events 
with all possible initial electron energies have been treated together and the average value of analysing power for the whole energy
range 200 -- 782~\keV\ was assumed to be $\left<S\right>=-0.18$.
\begin{figure}
  \center
  \includegraphics[width=0.8\textwidth]{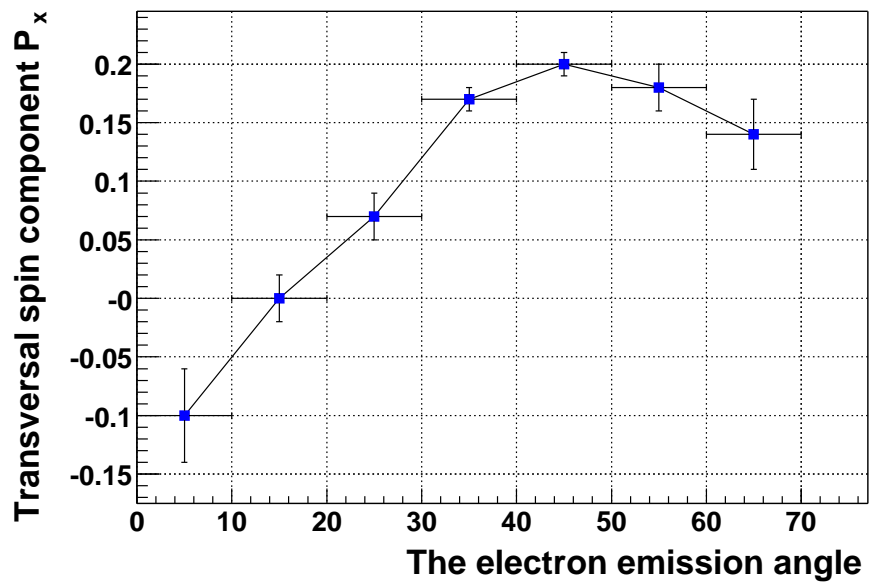}
  \caption{The false $R$ effect as a function of the electron emission angle.\label{fig-Rangles}}
\end{figure}
The result is shown in the Fig.~\ref{fig-Rangles}, one can see a clear dependency on the emission angle, what confirms that 
the nature of the systematic effect is well understood. For the emission angle below 50\degree\ its magnitude increases linearly
(a similar situation like in Ref.\cite{Li}), later it starts to decrease. It can be explained by the fact that for the larger 
emission angles electrons may scatter on the MWPC frame, therefore their final angle at the foil can be dramatically 
changed. Accurate estimation of the false contribution to the 
$R$ coefficient is essential for the final systematic uncertainty of the whole measurement. 
\cleardoublepage
\chapter{Summary and conclusions}
The main purpose of this thesis was the development of the comprehensive Monte Carlo simulation and the analysis of systematic 
uncertainties in the planned $R$-correlation coefficient measurement. The biggest effort
was required to create and test the simulation program, based on the \geant\ package and to develop appropriate
data analysis tools. In the first step, all previously written parts of software have been checked, debugged and 
unified in one application providing the full functionality. Furthermore, the new construction of the 
experimental apparatus had to be included replacing the old implementation. The most challenging part of the development
phase was, however, the necessity of creating the code for the simulation of polarised electron processes and for the 
polarisation transport. So far, such an option has not been provided by the authors of \geant\ package.

Finally, the program has been used to search for systematic effects in the measurement. The first effect associated
with the neutron $\beta$ decay asymmetry has been already recognized, explained and estimated. Due to the false contribution
to the $R$ coefficient induced by this phenomena, it is essential to control it as well as it is possible. The Monte Carlo
simulation provides the robust and powerful tool, which will serve for that purpose. Another important issue is the problem of 
the electron track reconstruction, which might introduce new systematic effects of different type. With artificial data, 
these contributions can be better controlled.

The search for other effects is in progress and definitely the program is going to be extensively used in future for the 
further measurement improvement and the data analysis. 
\cleardoublepage
\appendix
\chapter{The macro {\tt run.mac}}
\begin{verbatim}
/gun/PositionRandom on
/gun/PositionRandomType linear

/gun/EnergyRandom on
/gun/EnergyRandomType raw
/gun/Decay on
/gun/particle e-

/gun/MomentumRandom on
/gun/AngleMinus 70 degree
/gun/AnglePlus 70 degree

# Neutron beam properties
/gun/XBeam 45. mm
/gun/YBeam 145. mm
/gun/XDiv 0.8 degree
/gun/YuDiv 0.9 degree
/gun/YdDiv 0.65 degree
/gun/range 1.1
/gun/neutron_polarization 0.8973
/gun/NeutronSpinRandom on

# Neutron beta decay parameters
/gun/A 0.0
/gun/R 0.0
/gun/N 0.0

/chamber/threshold_v .4
/chamber/threshold_h .8
/chamber/sjit_v 10.
/chamber/sjit_h 10.
/chamber/rnoise_level_v 0.
/chamber/rnoise_level_h 0.
/chamber/effiplane 0.95

/run/beamOn 1
#/run/beamOn 50000000
\end{verbatim}
\cleardoublepage
\chapter{The macro {\tt geom.g4mac}}
\begin{verbatim}
#Titan Foil
#/mydet/boron_box/ti_foil_thickness 	5.393 micrometer 
# full thickness of the Ti foil (for energy loss simulation)
/mydet/boron_box/ti_foil_thickness 	0.0 micrometer 

# Boron box
/mydet/boron_box/bigger_x 		700. 	mm	# boron box width (with at the end of the box)
/mydet/boron_box/smaller_x 		214. 	mm	# boron box width (the central part)
/mydet/boron_box/height 		625. 	mm	# height of the b.b.
/mydet/boron_box/length 		2533.5 	mm	# total b.b. length
/mydet/boron_box/begin_length 		360. 	mm	# length of the beginning of the b.b.
/mydet/boron_box/end_length 		1278.5	mm	# length of the ending of the b.b.
/mydet/boron_box/border 		1.25 	mm	# total b.b. border thickness
/mydet/boron_box/mid_border 1.2 	mm	# thickness of the inner layer of the border

/mydet/boron_box/window/height 500. mm	# height of b.b. windows
/mydet/boron_box/window/width 500. mm	# width of b.b. windows
                                    # window placement:
/mydet/boron_box/window/y_pos 0. cm	# - height 
#(relative to the geometrical center of the b.b.)
/mydet/boron_box/window/z_pos 810. mm	# - z-axis position (along the beam)

# Dump
/mydet/boron_box/dump/width 650. mm 	# beam dump sizes
/mydet/boron_box/dump/height 622.5 mm 	#
/mydet/boron_box/dump/thickness 3.6 mm 	# and its placement
#/mydet/boron_box/dump/position 2530. mm # along the beam line

# MWPC
/mydet/mwpc/height 		625.	mm # MWPC sizes:
/mydet/mwpc/length 		625.	mm #	
/mydet/mwpc/thickness 		91. 	mm #
/mydet/mwpc/zposition_left      810. 	mm # rel. to the beam: left chamber
/mydet/mwpc/yposition_left      0. 	cm #                        
/mydet/mwpc/zposition_right     810. 	mm # right chamber
/mydet/mwpc/yposition_right     0. 	cm #                         

# Wire chamber
/mydet/mwpc/inside_height 		500. 	mm # chamber sizes (inside the MWPC)
/mydet/mwpc/inside_length 		500.  	mm #
/mydet/mwpc/anode_spacing 		5. 	mm # distance between anodes
/mydet/mwpc/dist_anode_cathode 		4. 	mm # distance between anode and cathode planes        
/mydet/mwpc/cathode_spacing 		2.5 	mm # distance between cathodes
/mydet/mwpc/anode_plane_spacing 	16. 	mm # distance between anode planes
/mydet/mwpc/anode_diameter   25. micrometer        # wire diameter (anodes)
/mydet/mwpc/cathode_diameter 25. micrometer        # wire diameter (cathodes)
/mydet/mwpc/dist_entr_anode 	        12. 	mm # distance between the chamber 
# entrance and the nearest anode
/mydet/mwpc/anode_planes 		0          # total planes number 

/mydet/mwpc/material/isobutane  	10.   	   # gas mixture: -percent of isobutane
/mydet/mwpc/material/methylal   	6.  	   #  	      -percent of methylal

# Scintillators
/mydet/scintillator/height 60. cm 	# scintillator sizes
/mydet/scintillator/width 10. cm 	# 
/mydet/scintillator/thickness 10. mm 	#
/mydet/scintillator/spacing 1. mm 	# scintillator spacing
/mydet/scintillator/number 6		# number of scintilators

# Golden (lead ?) foil
/mydet/foil/width 60. cm 	# foil sizes
/mydet/foil/height 60. cm 	#
/mydet/foil/foil in		# "in" or "out" (with and without the foil)

# Distances
#/mydet/to_left_mwpc	0. mm 		# between b.b. surface and the left MWPC window
/mydet/to_right_mwpc 	0. mm 		# between b.b. surface and the right MWPC window
/mydet/to_scintillator 	6.6	cm 	# between MWPC's and hodoscopes
/mydet/to_golden_foil 	3.2	cm 	# betweem MWPC's and the lead foil

# Nose (Collimator)  --------------------------------------  only rough values
/mydet/nose/bars_number 6	# number of barriers inside the collimator
/mydet/nose/to_barrier 50. mm 	# distance between the nose ending and 
# the first barrier
/mydet/nose/border 11.2 mm	# total thickness of the "nose" border
/mydet/nose/mid_border 1.2 mm	# thicknes of the inner layer of the border
\end{verbatim}

\end{document}